\documentclass[11pt]{article}
\usepackage[margin=1in]{geometry}
\usepackage{times,amsmath,amssymb,booktabs,xcolor}
\usepackage[hidelinks]{hyperref}
\usepackage{enumitem}
\usepackage{graphicx}
\usepackage{rotating}
\usepackage{tikz}\usetikzlibrary{arrows.meta,patterns}

\tikzset{every node/.style={font=\scriptsize}}

\newcommand{\Rey}{\mathcal{R}}

\newcommand{\Fhi}{F_{\rm hi}}
\newcommand{\OAI}{OpenAI~2026}
\newcommand{\GD}{GD1998}

\title{Self-similar swirl between contracting porous walls:\\
the \GD{} exact Navier--Stokes solution revisited in the similarity variables of the \OAI{} forced blow-up construction}
\newif\ifaiauthor\aiauthorfalse
\newcommand{\RD}{\ifaiauthor the first author\else the author\fi}\newcommand{\RDcap}{\ifaiauthor The first author\else The author\fi}
\providecommand{\doi}[1]{\href{https://doi.org/#1}{doi:#1}}   
\ifaiauthor
\author{Ramani Duraiswami\thanks{Department of Computer Science, University of Maryland, College Park; also a member of the Applied Mathematics and Scientific Computing Program. Email: \href{mailto:ramanid@umd.edu}{ramanid@umd.edu}. ORCID: \href{https://orcid.org/0000-0002-5596-8460}{0000-0002-5596-8460}.}
\and Claude Fable 5.1\thanks{Anthropic}}
\else
\author{Ramani Duraiswami\thanks{Department of Computer Science, University of Maryland, College Park; also a member of the Applied Mathematics and Scientific Computing Program. Email: \href{mailto:ramanid@umd.edu}{ramanid@umd.edu}. ORCID: \href{https://orcid.org/0000-0002-5596-8460}{0000-0002-5596-8460}.}}
\fi
\date{Preprint, 15 September 2026}

\begin{document}
\maketitle
\begin{center}\emph{In memory of \href{https://www.umiacs.umd.edu/news-events/news/nail-gumerov-1960-2022}{Nail A. Gumerov} (d.\ 2022), co-author of \GD{}, the 1998 work that this paper builds on.}\end{center}

\begin{abstract}
When the details of the OpenAI claim of finite-time blow-up for the forced Navier--Stokes equations (\OAI{} below) reached us on 12 September 2026, we wanted to check two things: whether the OpenAI construction is computable, and whether it can be related to a fluids experiment one could propose to verify it. Its object is an axisymmetric swirl core in cylindrical coordinates, written in anisotropic similarity variables. \RDcap{} was immediately struck by its similarity to work he had done with Nail Gumerov in 1998 (\GD{} below), where they had obtained an exact
steady solution for swirl between porous coaxial cylinders in the same coordinates. This was computed by Chebyshev collocation with continuation in the radial Reynolds number. We show that the \GD{} boundary value problem does not recast into the similarity variables, because the polynomial-in-$z$ closure fails, and that its generalization is a two-dimensional profile problem between porous walls held at fixed similarity radii, second order in the radial variable $X$ and first order in the time-like axial variable $\eta$. We
solve it by tensor-product Chebyshev collocation with an explicitly pinned pressure gauge, a complex-step Newton method and pseudo-arclength continuation, verify the discretization against symbolic derivations, an exact exterior solution and a manufactured solution, and sweep the radial Reynolds number $V_0\in[-50.9,50.1]$, the wall swirl $\Fhi\in[1,120]$ and a symmetry-breaking datum. On the full range of $\eta$ the solution is a single smooth branch for inflow below about $9$; above it the symmetric branch is an $S$-curve in swirl whose upper fold, at $(\Fhi,V_0)=(40.47,-10)$, is converged on three grids, and whose
returning sheet is not converged in $\eta$ on any grid tried. Two results of method govern these statements: cutting the $\eta$ range turns an outflow boundary into an inflow boundary once $|U|$ exceeds $D\eta_c/(1-\eta_c^2)$ at the cut and manufactures a spurious bifurcation structure; and an inner-wall layer at strong inflow that 64 plain radial modes do not resolve and 32 mapped ones do. Deflated Newton shows that the Dirichlet axis problem has no resolution-stable solution, which fixes the formulation of the blow-up core as a Cauchy problem in $X$, for which we give the recursion, the series and the march. Imposing the moment identities of the \OAI{} construction on that core shows that they cannot be met by a core symmetric about the dividing plane: the blow-up core is an axial through-flow, as its authors chose, and a non-symmetric core with free annulus content meets the identities to $0.2\%$ at the smallest exterior amplitude tried, with an axis pressure deficit of $0.34$--$0.61$ times the peak swirl velocity squared. For the second question we
work out what a real fluid does with the singularity: the blow-up is energetically free, the anomalous factor $\tau^{-h}$ is $1.4$ at $\tau=10^{-15}$, and the first continuum assumption to fail is cavitation in a liquid and compressibility in a gas, both while $\tau^{-h}$ is within $12\%$ of unity. Dynamic rescaling shows the profile to be an attractor of the collapse at weak inflow and along the approach to the fold, which is a saddle-node; at moderate inflow and strong swirl the spectrum does not converge with the axial resolution, the wandering modes living on a sonic line of the axial transport next to the outflow boundary, and the stability question there is left open. A porous-wall swirl chamber of the \GD{} type driven toward collapse is proposed as the experiment, with axis cavitation inception and the loss of the steady state at the fold as its two observables. Nothing found suggests that the mechanism is reachable in a flow one computes or builds, and the forced theorem says nothing about the unforced equations of engineering practice, which this study leaves as it found them. The solver, its tests, the reports and the research log accompany the paper; the open items are listed with the step that would settle each.
\end{abstract}

\begin{center}\small\begin{minipage}{0.92\textwidth}
\textbf{Status.} This is the record of a four-day study, 12 to 15 September 2026, and it is complete as it stands: the author is not continuing it. Everything stated as a result has been computed and checked as described in Section~\ref{sec:verif} and in the tests that accompany the code. The open items are stated where they arise and collected in Section~\ref{sec:disc}, each with the step that would settle it. The solver, its tests, the reports and the research log are at \url{https://gitlab.umiacs.umd.edu/ramanid/swirl-collapse} and accompany the paper as ancillary files.

\medskip
\textbf{Use of generative AI.} This work was carried out between 12 and 15 September 2026 by \RD{} working with Claude Fable 5.1 (Anthropic), a generative AI system used through Claude Code. Claude wrote and tested the solver and analysis code under \RD's direction, ran the computations on a laptop and on the Zaratan and Nexus clusters, kept the research log, and drafted and revised the text of this paper; \RD{} set the questions, supplied the 1998 work and its method, checked the derivations against the source paper, made the scientific judgments and edited the text. \ifaiauthor Both authors take responsibility for the contents. \else The author takes full responsibility for all contents. Claude is named here, rather than as an author, in accordance with arXiv's policy on generative AI. \fi
\end{minipage}\end{center}

\section{Introduction}\label{sec:intro}
\begin{figure}[!b]\centering
\includegraphics[width=0.62\textwidth]{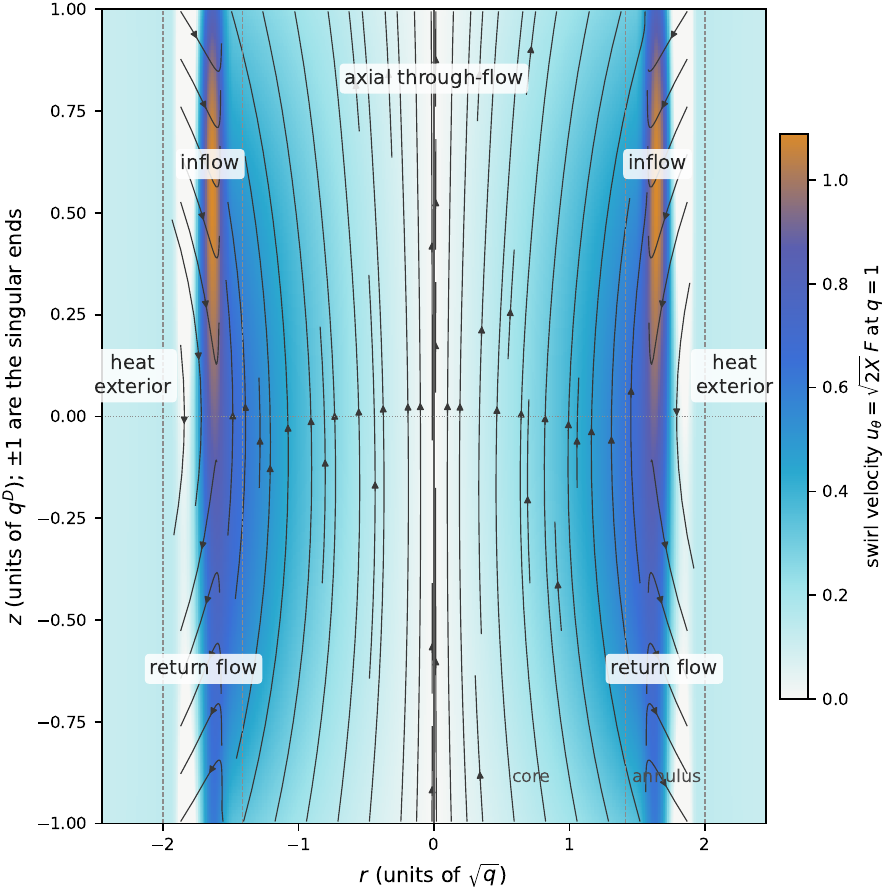}
\caption{The blow-up flow as computed: the leading-order profile of this work that meets the moment identities of \OAI{} at $c_\infty=0.2$ (Section~\ref{sec:axis}), in the meridional plane at the reference instant $q=1$. Color is the swirl velocity $u_\theta=\sqrt{2X}F$, the lines are streamlines of the meridional velocity $(u_r,u_z)$, the axis is at $r=0$, the dividing plane at $z=0$, the singular ends at $z=\pm1$; dashed lines mark the core, the forced annulus and the heat exterior, which carries swirl only. The three motions of the construction appear, inflow, axial through-flow along the axis and return flow at larger radius, but at this amplitude the circulation Reynolds number is of order one, so a fluid particle turns through a fraction of a revolution per decade of time to the singularity: the collapse is of the profile, not a winding of material lines.}\label{fig:hero}
\end{figure}

On 8 September 2026 a construction of finite-time blow-up for the three-dimensional Navier--Stokes equations with a smooth, compactly supported force was released by OpenAI \cite{openai26}; we refer to it as \OAI{} throughout. When the claim came up we wanted to check two things: (a) whether the \OAI{} construction is computable, that is, whether its leading-order profile can be produced as a number on a grid and its force evaluated; and (b) whether it can be related to an actual fluids experiment one could propose to verify it. Everything in this paper serves one of the two; Figure~\ref{fig:hero} shows the profile computed here.

The claim was announced on 8 September 2026 and drew wide attention that week, but \RD{} did not see its details then. On the morning of Saturday 12 September \RD{} read the \OAI{} construction, saw that its object is a vortical flow, a swirl core in cylindrical coordinates, and recognized the geometry of the porous-cylinder problem he had solved with Gumerov in 1998 \cite{gd98}, referred to as \GD{} throughout. The work reported here began that day; the recast, the solver, its verification, the sweep, the axis core, the estimates and the stability analysis were done between 12 and 15 September 2026, by \RD{} working with Claude as described in the statement on generative AI on page~2. Navier--Stokes is not \RD's research area today; this paper reports what someone returning to a problem left in 1998, with a validated solver and an AI collaborator, could establish in four days, and it ends with the questions that remain, stated so that others can take them up. The short answer to both questions is in Section~\ref{sec:disc}: the profile can be computed and its core is what the authors of \OAI{} chose; the annulus that sustains it lives at radii no computation reaches; a real fluid leaves the equations' description long before the singularity; and nothing here bears on the unforced equations that engineering practice solves, which the forced theorem also leaves untouched.

The object of the \OAI{} construction is an axisymmetric swirl core, spun up by inward radial flow and evacuated axially, written in anisotropic similarity variables. Its geometry is that of a flow we know: in 1998 \RD{} and Gumerov found an exact steady solution for swirl between two rotating porous coaxial cylinders with prescribed filtration through the walls, motivated by a patented microfiltration device \cite{gd98,chahine96}; the flow reduces to a nonlinear boundary value problem for one fourth-order ordinary differential equation,
which they solved by Chebyshev collocation with relaxation in the radial Reynolds number $\Rey$. Gol'dshtik and Ersh \cite{ge91,ge92} had treated the case without the inner cylinder by shooting and used it for stability studies of pipe flow with suction. Inward spiral, angular-momentum transport by the inflow, axial outflow, and a symmetry-breaking parameter that controls the axial velocity on the dividing plane are common to both.

For question (a) the \GD{} problem is a testbed, not the object. The \OAI{} theorem concerns an axis core joined to a heat-equation exterior through an annulus in which an oscillatory force supplies the momentum the leading-order profile lacks; nothing in it concerns wall-bounded flow with prescribed transpiration. What the porous annulus offers is a well-posed problem in the same variables, with the same leading-order operators, on which a solver for the core can be validated and on which the pitfalls of those operators can be found before they are met on the axis. We found two, a domain rule and a wall layer, and one negative result that fixes the formulation of the core. Along the way the steady \GD{} boundary value problem turns out not to ``recast'' into the similarity variables: the closure that made it one-dimensional fails, and what replaces it is a two-dimensional profile problem in the radial similarity variable $X$ and the axial one $\eta$, second order in $X$ and first order in $\eta$, with $\eta$ playing the role of time. For question (b) the similarity scalings alone say what a real fluid would do with the singularity and which continuum
assumption breaks first; the porous-wall chamber of \GD{} is the natural apparatus, and the sweep supplies its second observable.

\begin{figure}[!thb]\centering
\includegraphics[width=\textwidth]{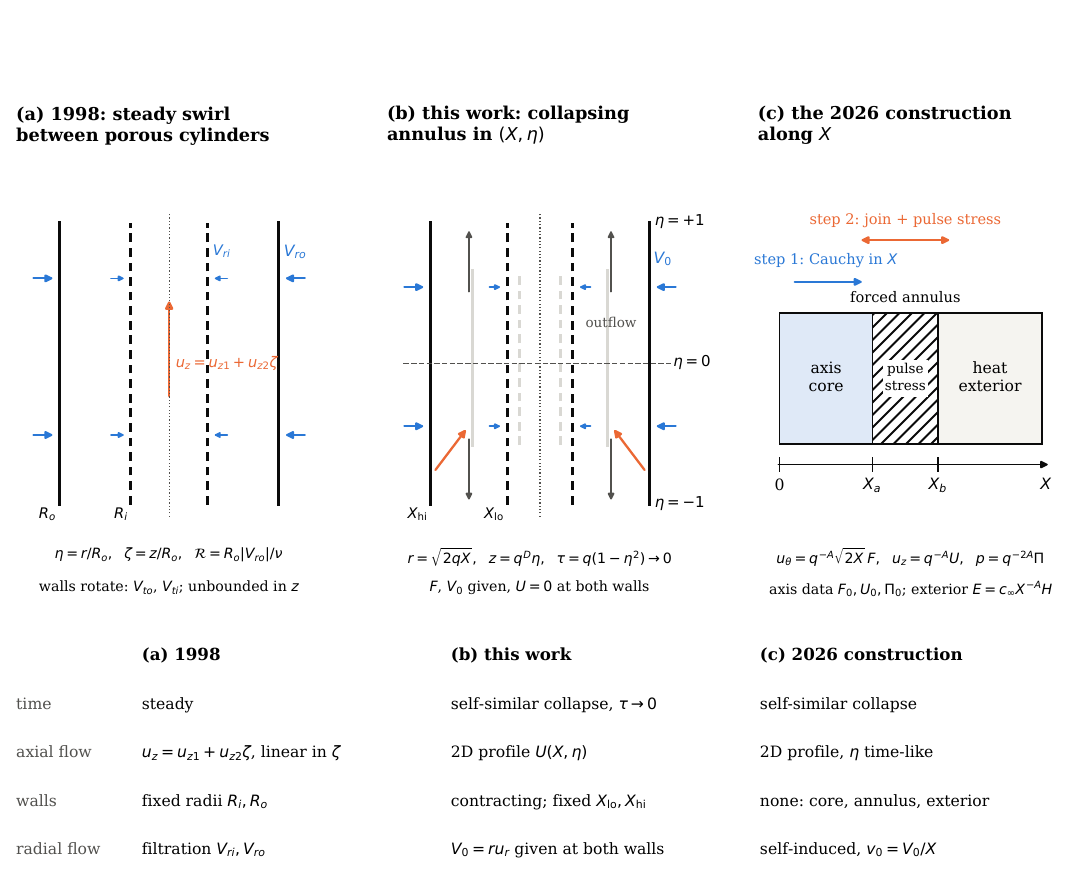}
\caption{The three flow scenarios. (a) The \GD{} configuration: steady swirl between two rotating porous cylinders with radial filtration through both walls and an axial flow linear in $\zeta$, unbounded in $z$. (b) The setting of this work: a porous annulus held at fixed similarity radii $X_{\rm lo},X_{\rm hi}$ whose physical radii shrink like $\sqrt{2qX}$ as $\tau\to0$; $\eta$ is time-like, with the dividing plane $\eta=0$
in the middle and outflow at $\eta=\pm1$, so no boundary data are imposed in $\eta$; the lighter copy is the same annulus at a later time. (c) The object of the \OAI{} theorem: an axis core determined by analytic axis data, a forced annulus in which the pulses supply the stress that the leading-order residual demands, and the heat exterior. The strip below lists the differences that matter for the numerics.}\label{fig:scenarios}
\end{figure}

The remainder of the paper is organized as follows. Section~\ref{sec:2026} states the \OAI{} result, introduces its similarity variables and profile variables with the physical names we use for them, writes down the leading-order core system, and says which of its branches we pursue and which we do not. Section~\ref{sec:1998} reproduces the \GD{} formulation and solution and connects each of its elements to those branches, ending with the dictionary between the two problems and the reason the \GD{} problem does not recast. Section~\ref{sec:num} describes the collocation, the pressure gauge, the Newton method and the continuation. Section~\ref{sec:verif} reports the verification against symbolic derivations, an exact exterior solution, a manufactured solution and a resolution study. Section~\ref{sec:sweep} reports the parameter sweep in the order in which its lessons were learned: the domain rule, the results on the full range, the wall layer at strong inflow, the fold locus and the stability of the profile under the collapse. Section~\ref{sec:axis} reports the negative result for the Dirichlet axis problem, the Cauchy-in-$X$ formulation that replaces it and the results of its series, march, join and matching stages. Section~\ref{sec:fluid} works out what a fluid would do with the singularity and proposes the experiment. Section~\ref{sec:disc} concludes with what is established, what is open, and where the material is; Appendix~\ref{app:methods} introduces each numerical method briefly and points to the code. Figure~\ref{fig:scenarios} sets the three flow scenarios side by side; Tables~\ref{tab:methods} and~\ref{tab:params} summarize the methods and the parameter ranges.

\section{The \OAI{} construction, recast}\label{sec:2026}
\paragraph{The result.} Theorem 1.1 of \OAI{} \cite{openai26} states that for every $\nu>0$ there exist a force $f\in C_c^\infty(\mathbb R^3\times(0,\infty))$ and a smooth solution $(u,p)$ on $\mathbb R^3\times[0,1)$ with $u(\cdot,0)=0$, compact spatial support, $\sup_t\|u\|_{L^2}<\infty$ and $\limsup_{t\uparrow1}\|u\|_{L^\infty}=\infty$. This addresses alternatives (C) and (D) of the Clay problem statement, breakdown with a smooth compactly supported force; the unforced alternatives are untouched. The proof was produced by an automated system and is accompanied by a Lean~4 formalization; as of 12 September 2026 there was no refereed independent verification. Nothing in the theorem concerns wall-bounded flow.

\paragraph{Similarity variables and the leading-order core.} With $\nu=1$, cylindrical coordinates $(r,\theta,z)$ and axisymmetric fields, the \OAI{} construction uses
\begin{gather}
\tau=1-t,\qquad A=\tfrac12+h,\qquad D=\tfrac12-h,\qquad z=q^D\eta,\qquad \tau=q(1-\eta^2),\notag\\
X=\frac{r^2}{2q},\qquad d=1-\eta^2,\qquad L=1-2h\eta^2,
\end{gather}
with $0<h<1/100$. We give these quantities physical names and use the names alongside the symbols. $\tau=1-t$ is the time to the singularity; $q$ is the collapse scale, the shrinking length-squared scale of the core, equal to $\tau$ on the plane $z=0$; $X=r^2/2q$ is the viscous similarity radius, the squared radius measured in units of the collapse scale; $\eta=z/q^D$ is the axial similarity coordinate, and it is time-like, since at fixed collapse scale $\eta\to\pm1$ means $\tau\to0$: $\eta=0$ is the dividing plane and $\eta=\pm1$ are the singular ends. $h$ is the anisotropy exponent and $A$, $D$ the collapse exponents of velocity and height; $d=1-\eta^2$ and $L=1-2h\eta^2$ are the end factors, both equal to one on the dividing plane and $d$ vanishing at the singular ends. The smooth profile variables are
\begin{equation}\label{eq:smooth}
u_\theta=q^{-A}\sqrt{2X}\,F,\qquad u_z=q^{-A}U,\qquad r\,u_r=X v_0,\qquad p=q^{-2A}\Pi .
\end{equation}
$F$ is the reduced swirl, smooth on the axis; $E=\sqrt{2X}\,F$ is the swirl profile and $\mathcal H=2XF$ the angular-momentum profile ($ru_\theta=q^{-A}\sqrt q\,\mathcal H$); $U$ is the axial profile; $V_0=Xv_0=ru_r$ is the wall inflow, which with $\nu=1$ is the radial Reynolds number, and $v_0$ its reduced form; $\Pi$ is the similarity pressure and its value on the inner boundary, $\Pi_0(\eta)$, the pressure datum. For any $b$ the time and axial derivatives act on $q^bf$ through the operators
\begin{align}\label{eq:TZ}
\partial_t(q^bf)&=q^{b-1}T_bf, & T_bf&=L^{-1}(-bf+D\eta f_\eta+Xf_X),\notag\\
\partial_z(q^bf)&=q^{b-D}Z_bf, & Z_bf&=L^{-1}(2b\eta f+df_\eta-2\eta Xf_X),
\end{align}
which we verified symbolically (Section~\ref{sec:verif}). Substituting \eqref{eq:smooth} into the
axisymmetric Navier--Stokes equations and collecting powers of $q$, the leading balance retains radial
viscosity and drops axial viscosity and radial inertia, which are $O(q^{2h})$ smaller:
\begin{subequations}\label{eq:core}
\begin{align}
&T_{-(A+\frac12)}F+v_0\,(XF_X+F)+U\,Z_{-(A+\frac12)}F-2\,(XF)_{XX}=0, &&\text{[$\theta$]}\label{eq:th}\\
&T_{-A}U+Xv_0\,U_X+U\,Z_{-A}U+Z_{-2A}\Pi-2\,(XU_X)_X=0, &&\text{[$z$]}\label{eq:z}\\
&(Xv_0)_X=L^{-1}\big(2A\eta U-dU_\eta+2\eta XU_X\big), &&\text{[cont.]}\label{eq:cont}\\
&\Pi_X=F^2 . &&\text{[rad.]}\label{eq:rad}
\end{align}
\end{subequations}
Equations \eqref{eq:cont}--\eqref{eq:rad} are eqs.~(4.7) of \OAI; vanishing of the residuals of
\eqref{eq:th}--\eqref{eq:z} is its inner-region system (4.13), and in the pulse annulus those residuals
are instead the divergence of the stress the oscillatory force supplies. The pressure datum
$\Pi_0(\eta):=\Pi(X_{\rm lo},\eta)$ is a free function, the analogue of the constants $(\alpha,\beta)$ in
\eqref{eq:ansatz98}.

The system is second order in the viscous similarity radius $X$ and first order in the axial similarity coordinate $\eta$. The coefficient of $F_\eta$ in \eqref{eq:th}
is $L^{-1}(D\eta+dU)$; at the singular ends $\eta=\pm1$ it equals $\pm D/L$, so both ends are outflow and no boundary
condition may be imposed there, while characteristics emanate from the curve $D\eta+dU=0$ near the
dividing plane $\eta=0$. The sign of the axial profile $U(X,0)$ on the dividing plane decides which way information leaves it; it is the
\OAI{} counterpart of the \GD{} parameter $\beta-G$ of Section~\ref{sec:1998}.

\paragraph{The heat exterior.} Outside the core, where $U$ and $v_0$ are negligible, \eqref{eq:th} reduces to a linear equation for the swirl alone, $T_{-(A+1/2)}F-2(XF)_{XX}=0$, which is the azimuthal heat equation $\partial_tu_\theta=\nu(\Delta-r^{-2})u_\theta$ of a decaying line vortex written in the similarity variables. \OAI{} solves it exactly: $E=\sqrt{2X}F=c_\infty X^{-A}H(2d/X)$ with $H(Z)=\Gamma(1+h)^{-1}\int_0^\infty e^{-v}v^h(1+Zv)^{-h}\,dv$, and this heat exterior is the far field into which the forced annulus feeds the core. It has no counterpart in \GD{}, where the outer porous wall closed the domain, and its addition to our problem plays five roles below. It is the exact solution against which the discretization is verified (Section~\ref{sec:verif}). Its swirl tail $u_\theta\sim r^{-1-2h}$ is what the \GD{} tail is matched to, giving $\Rey=2+2h$ (Section~\ref{sec:1998}). It is smooth but not analytic at the singular ends $\eta=\pm1$, and this alone makes the convergence in $\eta$ algebraic on the full range (Sections~\ref{sec:verif} and~\ref{sec:sweep}). In the axis problem it is the target of the join: its amplitude $c_\infty$ is the one normalization the moment identities do not fix, its pressure $\Pi_{\rm ext}=-\int_X^\infty F_{\rm ext}^2\,dx$ fixes the axis pressure datum, and its tail $c_\infty^2X_b^{-2h}/(4h)$ in the identity $S(\infty,\eta)=0$ is the term that forces the axial through-flow (Section~\ref{sec:axis}). And because $H(Z)\sim Z^{-h}$ at large $Z$, toward the axis it behaves as the potential vortex $F\sim0.706/X$, so core and exterior are not analytic continuations of each other and must meet in the annulus, where the pulses act.

\paragraph{Which branches we pursue.} The \OAI{} construction has several parts, and we take up four of them. (i) The leading-order inner system in profile variables, eq.~(4.13) of \OAI, solved as a boundary-value problem between porous walls held at fixed viscous similarity radii: this is the sweep of Section~\ref{sec:sweep}. (ii) The axis core as the Cauchy problem in $X$ of Proposition B.2 of \OAI, series and march (Section~\ref{sec:axis}). (iii) The join of that core to the heat exterior with the stress of Section~3.2 of \OAI, the moment identities of its Theorem~4.6(v) and the cone condition of its Section~4.3, implemented and tested: the identities force an axial through-flow on the dividing plane and are met to $0.2\%$ by a non-symmetric core with free annulus content at the smallest exterior amplitude tried, while the cone fails on every smooth profile (Section~\ref{sec:axis}). (iv) The estimates of where a real fluid arrests the collapse (Section~\ref{sec:fluid}). We do not pursue the pulse construction itself, the radial oscillation with phase $N\log X$ of Propositions~7.5 and C.2 of \OAI by which the annulus stress is realized as a smooth force; nor the higher-order corrections in $q^{2h}$ to the leading-order system; nor the full forced Navier--Stokes evolution at finite viscosity, which is a computation of a different scale.

\section{The \GD{} solution}\label{sec:1998}
Two vertical porous coaxial cylinders of radii $R_i<R_o$ rotate with tangential velocities $V_{ti}$, $V_{to}$; liquid filters through them with radial velocities $V_{ri}$, $V_{ro}$ and enters and leaves through the annuli at the bottom and top. This porous-wall geometry is the annulus of branch (i) of Section~\ref{sec:2026}, with the walls held at fixed radii rather than at fixed viscous similarity radii, and the chamber built on it at Dynaflow is the apparatus of branch (iv). With $\eta=r/R_o$, $\zeta=z/R_o$, $\eta_i=R_i/R_o$, velocities scaled by a reference $V_*$, $\Rey=R_oV_*/\nu_l$, $P=(p-p_*)/(\rho_lV_*^2)$ and $G=R_og/V_*^2$, \GD{} sought steady axisymmetric solutions of the form 
\begin{equation}\label{eq:ansatz98}
u_r=u_r(\eta),\qquad u_z=u_{z1}(\eta)+u_{z2}(\eta)\,\zeta,\qquad u_\theta=u_\theta(\eta),\qquad
P=\Pi(\eta)-\tfrac12\alpha\zeta^2-\beta\zeta .
\end{equation}
Separation of the variables $\eta$ and $\zeta$ in the Navier--Stokes equations gives
\begin{align}
\frac1\eta(\eta u_r)'+u_{z2}&=0,\label{eq:98cont}\\
u_ru_r'-\frac{u_\theta^2}{\eta}+\Pi'&=\Rey^{-1}\Big(\frac1\eta(\eta u_r)'\Big)',\label{eq:98r}\\
u_ru_\theta'+\frac{u_ru_\theta}{\eta}&=\Rey^{-1}\Big(\frac1\eta(\eta u_\theta)'\Big)',\label{eq:98th}\\
u_ru_{z1}'+u_{z1}u_{z2}&=(\beta-G)+\frac{\Rey^{-1}}{\eta}(\eta u_{z1}')',\label{eq:98z1}\\
u_ru_{z2}'+u_{z2}^2&=\alpha+\frac{\Rey^{-1}}{\eta}(\eta u_{z2}')',\label{eq:98z2}
\end{align}
with $u_r=u_{ri}$, $u_\theta=u_{ti}$, $u_{z1}=u_{z2}=0$ at $\eta=\eta_i$ and $u_r=u_{ro}$,
$u_\theta=u_{to}$, $u_{z1}=u_{z2}=0$ at $\eta=1$. Equations \eqref{eq:98cont} and \eqref{eq:98z2} close on
$(u_r,u_{z2})$; eliminating $u_{z2}$ and the constant $\alpha$ by one differentiation gives a fourth-order equation for the radial velocity alone,
\begin{equation}\label{eq:98four}
u_r^{IV}+\frac2\eta u_r'''-\frac3{\eta^2}u_r''+\frac3{\eta^3}u_r'-\frac3{\eta^4}u_r
=\Rey\Big[u_ru_r'''-u_r'u_r''-\frac1\eta u_ru_r''-\frac1\eta(u_r')^2-\frac3{\eta^2}u_ru_r'+\frac4{\eta^3}u_r^2\Big],
\end{equation}
subject to $u_r=u_{ri}$, $u_r'+u_r/\eta_i=0$ at $\eta_i$ and $u_r=u_{ro}$, $u_r'+u_r=0$ at $1$. Given $u_r$, equations \eqref{eq:98z1} and \eqref{eq:98th} are linear in $u_{z1}$ and $u_\theta$, and $\Pi$ follows from \eqref{eq:98r} up to a constant. The parameter $\Rey$ controls the strength of the nonlinearity; the combination $\beta-G$ is the axial pressure gradient in excess of hydrostatic, and \begin{equation}\label{eq:98sym}
\beta=G\quad\Longrightarrow\quad u_{z1}\equiv0,\qquad u_z\big|_{z=0}=0,
\end{equation}
which is the reflection-symmetric case used by Gol'dshtik and Ersh; $\beta=G$ is the \GD{} form of the reflection symmetry $U(X,0)=0$ about the dividing plane that organizes branch (i). The \OAI{} core is not of this kind: its Theorem~4.6(v) forces $U(X,0)\neq0$ (Section~\ref{sec:axis}), so the case $\beta\neq G$ that \GD{} treated as a perturbation is the one the \OAI{} construction lives in.

\paragraph{Closed form.} When $u_{z2}\equiv0$ and $\alpha=0$, \eqref{eq:98cont} integrates to $u_r=c\eta^{-1}$, which is admissible only if $u_{ro}=u_{ri}\eta_i=c$; with $V_*=|V_{ro}|$ one has $\Rey=R_o|V_{ro}|/\nu_l=R_i|V_{ri}|/\nu_l$ and $c=\operatorname{sgn}V_{ro}=\pm1$, and
\begin{align}
u_{z1}&=\frac{\Rey(G-\beta)(\eta_i^2-1)}{2(2-c\Rey)}\left(\frac{\eta^2-1}{\eta_i^2-1}-\frac{\eta^{c\Rey}-1}{\eta_i^{c\Rey}-1}\right),\label{eq:98uz1}\\
u_\theta&=\frac1\eta\left[u_{to}+(u_{ti}-u_{to})\frac{\eta^{2+c\Rey}-1}{\eta_i^{2+c\Rey}-1}\right].\label{eq:98uth}
\end{align}
As $V_{ro}\to0$ this recovers the flow between rotating coaxial cylinders; at $c\Rey=2$ the expressions degenerate and their limit contains logarithms. The swirl tail $u_\theta\sim\eta^{1+c\Rey}$ and its degeneracy at $|c\Rey|=2$ are the \GD{} counterpart of the exterior tail match of the \OAI{} construction, which fixes $\Rey=2+2h$ (see the dictionary below). Everything in this section is from \cite{gd98}.

\paragraph{Method.} \GD{} discretized \eqref{eq:98four} by Chebyshev collocation on Gauss--Lobatto points \cite{boyd,canuto}, collocating the equation at the interior points and the boundary conditions at the end points, and solved the nonlinearity by successive substitution: the linear $\Rey=0$ problem first, then the nonlinear right-hand side treated as known with the linear operator LU-factored once, with $\Rey$ increased gradually to its target; this relaxation in $\Rey$ is the ancestor of the pseudo-arclength continuation of Section~\ref{sec:num}, which replaces it because it cannot pass a fold. Collocation was preferred to the shooting of Gol'dshtik and Ersh because their minimization over several shooting parameters was ill-posed. The solver was validated against \eqref{eq:98uz1}--\eqref{eq:98uth} over $\Rey\in[0,20]$ and tangential Reynolds numbers to $1000$; the worked example was $\eta_i=0.2$, $\Rey=1$, $\beta-G=0$, $u_{to}=50$, $u_{ro}=u_{ri}=-1$, $u_{ti}=0$, and particle paths in that flow were integrated with a fourth-order Runge--Kutta method.

\paragraph{Dictionary.} With both formulations in hand: the \GD{} radius $\eta_{98}=r/R_o$ corresponds to $\sqrt{2X}$, the square root of twice the viscous similarity radius, up to the shrinking scale $\sqrt q$;
$\zeta$ to the axial similarity coordinate $\eta$, the linear ansatz $u_z=u_{z1}+u_{z2}\zeta$ being the two-term Taylor expansion of
$U(X,\eta)$ about $\eta=0$ with $u_{z1}\leftrightarrow U(X,0)$ and $u_{z2}\leftrightarrow U_\eta(X,0)$;
$ru_r=c$ of \eqref{eq:98uz1} to the wall inflow $V_0=Xv_0$, so that the \GD{} radial Reynolds number is $O(|V_0|)$ and the
$\mathrm{Re}_r=O(1)$ of \OAI{} is the \GD{} regime $\Rey=O(1)$; $u_{to}$ to the reduced swirl $F$ at the outer wall. Matching
the \GD{} swirl tail $\eta^{1+c\Rey}$ to the exterior $u_\theta\sim r^{-1-2h}$ of the \OAI{} construction requires
$\Rey=2+2h$ with inflow, which as $h\to0$ lands on the logarithmic degeneracy $|c\Rey|=2$ of
\eqref{eq:98uz1}--\eqref{eq:98uth}. The match is suggestive only: the \GD{} tail is a steady
advection--diffusion balance for the circulation, the \OAI{} one unsteady self-similar diffusion.

\paragraph{Why the \GD{} problem does not recast.} One cannot substitute the ansatz \eqref{eq:ansatz98} into
\eqref{eq:core}: the operators \eqref{eq:TZ} carry the $\eta$-dependent coefficients $d$ and $L$, so a
profile polynomial in $\eta$ does not close. The correct generalization is the two-dimensional problem
\eqref{eq:core} on an annulus $X_{\rm lo}\le X\le X_{\rm hi}$ between porous walls held at fixed similarity
radii (physical radii shrinking like $\sqrt{2qX}$), with prescribed swirl $F$ and flux $V_0$ at both walls
and $U=0$ there, as in the \GD{} boundary conditions. The question then is whether a self-similar collapsing
swirl exists between such walls; Section~\ref{sec:sweep} answers it.

\section{Numerical method}\label{sec:num}
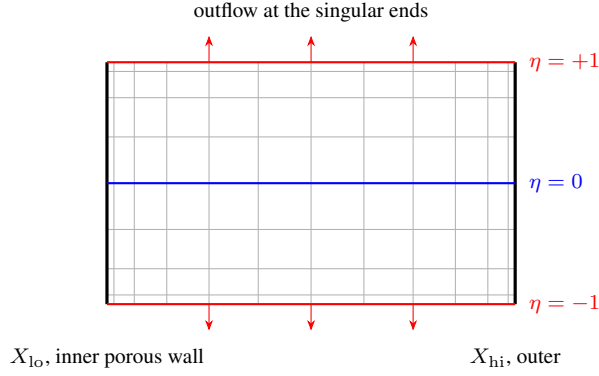
\begin{figure}[!t]\centering
\begin{tikzpicture}[x=1cm,y=1cm]
\foreach \i in {0,...,12}{\pgfmathsetmacro{\xx}{5.4*(1-cos(180*\i/12))/2}\draw[gray!60,thin] (\xx,0)--(\xx,3.2);}
\foreach \j in {0,...,8}{\pgfmathsetmacro{\yy}{1.6*(1-cos(180*\j/8))}\draw[gray!60,thin] (0,\yy)--(5.4,\yy);}
\draw[very thick] (0,0)--(0,3.2); \draw[very thick] (5.4,0)--(5.4,3.2);
\draw[thick,blue] (0,1.6)--(5.4,1.6);
\draw[thick,red] (0,0)--(5.4,0); \draw[thick,red] (0,3.2)--(5.4,3.2);
\foreach \xx in {1.35,2.7,4.05}{\draw[-{Stealth[length=4pt]},red] (\xx,3.2)--(\xx,3.55); \draw[-{Stealth[length=4pt]},red] (\xx,0)--(\xx,-0.35);}
\node[anchor=north] at (0,-0.45) {$X_{\rm lo}$, inner porous wall};
\node[anchor=north] at (5.4,-0.45) {$X_{\rm hi}$, outer};
\node[anchor=west,blue] at (5.45,1.6) {$\eta=0$};
\node[anchor=west,red] at (5.45,3.2) {$\eta=+1$};
\node[anchor=west,red] at (5.45,0) {$\eta=-1$};
\node[anchor=south] at (2.7,3.6) {outflow at the singular ends};
\end{tikzpicture}
\caption{The tensor grid of the annulus problem in the viscous similarity radius $X$ (horizontal, $12$ Chebyshev nodes shown) and the axial similarity coordinate $\eta$ (vertical, $8$ nodes), clustered at the walls and at the singular ends. The dividing plane $\eta=0$ (blue) carries the gauge pin; the singular ends $\eta=\pm1$ (red) are outflow and receive no boundary data.}\label{fig:gridnum}
\end{figure}
We use tensor-product Chebyshev--Gauss--Lobatto collocation on $N_X\times N_\eta$ nodes (Figure~\ref{fig:gridnum}) with
differentiation matrices in each direction and spectral integration in $X$ for $V_0$ and $\Pi$
\cite{trefethen00}. The unknowns are the reduced swirl $F$ and the axial profile $U$ at all nodes and the pressure datum $\Pi_0$ at the $\eta$ nodes when the wall inflow
is prescribed at both walls. Equations \eqref{eq:th}--\eqref{eq:z} are collocated at the interior $X$ nodes
and at all $\eta$ nodes including the ends, the polynomial ansatz selecting the smooth branch as it does
for a regular-singular ordinary differential equation collocated at its singular point; Dirichlet data for
$F,U$ are imposed at the walls and the flux $v_0=V_0/X$ at the outer wall.

\paragraph{Pressure gauge.} The discrete Jacobian has one exact null direction,
$\Pi_0\mapsto\Pi_0+\kappa(1-\eta^2)^{-2A}$, the physical gauge $p\mapsto p+\kappa\tau^{-2A}$. We fix it by
replacing the outer-flux collocation row at $\eta=0$ with $\Pi_0(0)=0$ and report the dropped row as a
compatibility residual. This requires an even $N_\eta$: the discrete gauge is an exact null vector only for
symmetric node sets containing $\eta=0$, where the collocated first-order gauge equation has an exact
even-polynomial solution; for odd $N_\eta$ the null vector is approximate ($\sigma\approx3\cdot10^{-11}$)
and pinning a flux row makes the reduced system singular.

\paragraph{Newton and continuation.} The Jacobian is formed by complex-step differentiation (exact to
roundoff, one residual evaluation per unknown) and the Newton system is solved by LU. Table~\ref{tab:cost}
gives the measured cost per Newton step on an 8-core laptop and the alternatives we tried; the rank-revealing
QR least-squares solve reproduces the SVD pseudo-inverse step to $10^{-12}$ at a quarter to a ninth of the
cost and served as the interim solver before the gauge was pinned. Continuation is Keller's pseudo-arclength
method \cite{keller77} with a bordered LU system, the arclength measured in the relative root-mean-square
norm of the solution so that the step is in parameter units where the solution changes slowly, folds detected
by the sign change of the parameter component of the tangent and refined by bisection, and simple branch
points flagged by a sign change of the bordered determinant without a fold. Symmetry-breaking points are
confirmed by the parity of the Jacobian's null vector (odd $F$, even $U$) and by the imperfect-bifurcation
test, a continuation in the asymmetry datum on either side of the suspected point.

\begin{table}[h]\centering\small
\begin{tabular}{lccccc}\toprule
grid & unknowns & complex-step Jacobian & SVD & RRQR (\texttt{gelsy}) & LU \\\midrule
$16\times8$  & 315  & 0.16\,s & 0.03\,s & 0.01\,s & 0.001\,s \\
$24\times12$ & 663  & 0.16\,s & 0.15\,s & 0.04\,s & 0.005\,s \\
$32\times16$ & 1139 & 0.38\,s & 1.20\,s & 0.14\,s & 0.016\,s \\
$40\times20$ & 1743 & $\sim$1\,s & 1.81\,s & 0.45\,s & 0.041\,s \\\bottomrule
\end{tabular}
\caption{Cost of one Newton step (best of three, 8-core laptop). With the gauge pinned the Jacobian is the
remaining cost; a warm-started continuation point takes two Newton iterations.}\label{tab:cost}
\end{table}

\begin{table}[!tp]\centering\small
\setlength{\tabcolsep}{4pt}
\setlength{\arrayrulewidth}{0.3pt}
\begin{tabular}{p{2.1cm}|p{3.0cm}|p{5.6cm}|p{4.3cm}}\toprule
component & \GD{} & this work & status \\\midrule
discretization & Chebyshev collocation in $\eta$ (1D) & tensor Chebyshev--Gauss--Lobatto in $(X,\eta)$, even $N_\eta$ (the discrete gauge is an exact null vector only for node sets containing $\eta=0$); sinh-mapped $X$ grid clustered at the inner wall for strong inflow & works; $\eta$ convergence on the full range is algebraic (open) \\
nonlinear solve & successive substitution, relaxation in $\Rey$ & Newton, complex-step Jacobian (exact to roundoff, $0.2$--$1$~s) & works \\
linear solve & LU & SVD pseudo-inverse $\to$ rank-revealing QR ($4$--$9\times$ faster, same step to $10^{-12}$) $\to$ gauge-pinned LU ($10$--$30\times$ faster again) & works \\
continuation & natural, in $\Rey$ & pseudo-arclength with solution-relative norm, fold bisection, det-sign branch flag & works; needs $ds_{\max}\le0.3$ at sharp folds (slow there) \\
bifurcation tests & none & null-vector parity, projection test (calibrated at three folds: $2$--$5\cdot10^{-3}$; $10^{-12}$ at a symmetry-breaking point), delta-line imperfect-bifurcation test, deflated Newton & works for symmetry breaking; projection test blunt at folds \\
$\eta$ domain & not present & full range required once $|U|>D\eta_c/(1-\eta_c^2)$; a cut admits inflow and produces spurious bifurcations & rule established; convergence in $\eta$ then algebraic (open) \\
strong inflow & $\Rey\le20$ & inner-wall layer, unresolved on plain grids to $N_X=64$; resolved by the mapped grid at $32\times16$ & works \\
axis problem & not present & Dirichlet form has no resolution-stable solution (deflation); Cauchy-in-$X$ core by series and filtered march, join to the heat exterior, pulse stresses, cone test and moment identities of \OAI{} implemented; identities closed to $0.2\%$ at $c_\infty=0.2$ & works to $c_\infty=0.2$; larger amplitude bounded by the fold of the pressure datum; cone open \\\bottomrule
\end{tabular}
\caption{The numerical methods of \GD{} and of this work, component by component, with the status found in the sweep. Timings are for an 8-core laptop.}\label{tab:methods}
\end{table}

\section{Verification}\label{sec:verif}
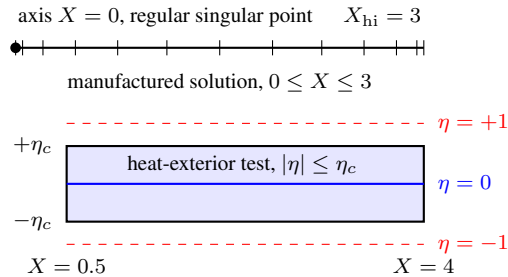
\begin{figure}[!t]\centering
\begin{tikzpicture}[x=1cm,y=1cm]
\draw[thick] (0,2.6)--(5.4,2.6);
\foreach \i in {0,...,12}{\pgfmathsetmacro{\xx}{5.4*(1-cos(180*\i/12))/2}\draw (\xx,2.5)--(\xx,2.7);}
\fill (0,2.6) circle (2pt);
\node[anchor=south west] at (-0.1,2.75) {axis $X=0$, regular singular point};
\node[anchor=south east] at (5.5,2.75) {$X_{\rm hi}=3$};
\node[anchor=north] at (2.7,2.4) {manufactured solution, $0\le X\le3$};
\draw[dashed,red] (0.675,0)--(5.4,0); \draw[dashed,red] (0.675,1.6)--(5.4,1.6);
\fill[blue!10] (0.675,0.3) rectangle (5.4,1.3); \draw[thick] (0.675,0.3) rectangle (5.4,1.3);
\draw[thick,blue] (0.675,0.8)--(5.4,0.8);
\node[anchor=west,red] at (5.45,1.6) {$\eta=+1$}; \node[anchor=west,red] at (5.45,0) {$\eta=-1$};
\node[anchor=east] at (0.62,1.3) {$+\eta_c$}; \node[anchor=east] at (0.62,0.3) {$-\eta_c$};
\node[anchor=west,blue] at (5.45,0.8) {$\eta=0$};
\node[anchor=north] at (0.675,-0.05) {$X=0.5$}; \node[anchor=north] at (5.4,-0.05) {$X=4$};
\node at (3.0,1.05) {heat-exterior test, $|\eta|\le\eta_c$};
\end{tikzpicture}
\caption{Domains of the two solution tests: the axis grid $0\le X\le3$ of the manufactured solution (top), with the regular singular point at $X=0$ collocated, and the strip $0.5\le X\le4$, $|\eta|\le\eta_c$ of the heat-exterior test (bottom), cut short of the singular ends where the exterior is not analytic.}\label{fig:gridverif}
\end{figure}
Every operator, the derivation, and the discretization were checked independently of the sweep (Figure~\ref{fig:gridverif} shows the two solution-test domains):
\begin{enumerate}[leftmargin=*]
\item Lemma 4.1, i.e.\ the operators \eqref{eq:TZ}, verified by brute-force chain rule with
\texttt{sympy} ($q_t=-1/L$, $q_z=2\eta q^{1-D}/L$); residual zero.
\item The system \eqref{eq:core} derived from the axisymmetric Navier--Stokes equations symbolically and
found identical to the coded residuals; the neglected terms carry $q^{2h}$ or $q^{4h}$.
\item The exact heat exterior $E=c_\infty X^{-A}H(2d/X)$, $H(Z)=\Gamma(1+h)^{-1}\int_0^\infty e^{-v}v^h(1+Zv)^{-h}dv$,
with $U=v_0=0$ solves \eqref{eq:th} (checked to $10^{-32}$ with \texttt{mpmath}). $H$ is $C^\infty$ but not
analytic at $Z=0$, i.e.\ at $\eta=\pm1$, so the collocation residual converges geometrically on
$|\eta|\le\eta_c<1$ and algebraically on $[-1,1]$: $\|R_\theta\|_\infty=6.9\cdot10^{-3},\,2.6\cdot10^{-3},\,
9.8\cdot10^{-4},\,3.7\cdot10^{-4}$ at $16\times8$ to $40\times20$ for $\eta_c=1$, against
$1.2\cdot10^{-4},\,4.9\cdot10^{-7},\,1.2\cdot10^{-8}$ for $\eta_c=0.6$ (the $10^{-8}$ floor being the
quadrature accuracy of $H$).
\item A manufactured solution on the axis domain $X\in[0,3]$: discrete and symbolic residuals agree to
$10^{-7}$, $8\cdot10^{-13}$ and $4\cdot10^{-12}$ at $N_X=12,20,28$.
\item The gauge-pinned LU solve reproduces the SVD pseudo-inverse solve to $1.4\cdot10^{-11}$ in $F$ and
$U$ and to $2\cdot10^{-11}$ in $\Pi_0$ modulo the discrete gauge; the arclength code locates the fold of
$y^2+\lambda-1=0$ to $10^{-9}$ and flags the branch point of $\lambda y-y^3=0$; symmetric data give
$|U(X,0)|=2\cdot10^{-15}$.
\end{enumerate}

\paragraph{Resolution at the reference point.} Six digits at $32\times16$ and seven at $40\times20$: Table~\ref{tab:line0} is the resolution study at the \GD{}
example with unit outer swirl ($V_0=-1$ at both walls, $F=1$ at the outer and $0$ at the inner wall,
$\eta_i=0.2$, $X_{\rm hi}=1$, $h=0.01$), comparing Chebyshev interpolants at fixed points. Convergence is
geometric at about a factor 20 per eight added modes per direction; $32\times16$ carries six digits and
$40\times20$ seven. At this point the interpolated values agree across $\eta_c=0.6,0.8,0.9$ to $10^{-11}$ in
$F$ and $10^{-10}$ in $U$, as the outflow structure requires (Section~\ref{sec:sweep} shows when this
stops being true). The smallest singular value of the gauge-fixed Jacobian decays spectrally with resolution
($6\cdot10^{-7}$, $9\cdot10^{-8}$, $6\cdot10^{-9}$, $4\cdot10^{-10}$, $3\cdot10^{-11}$ at $16\times8$ to
$48\times24$, $\eta_c=0.9$), a property of the continuous problem rather than of the discretization: the
$\eta$-direction has no boundary condition and the problem is unique only in the analytic class, so
polynomials of increasing degree approximate the smooth non-analytic homogeneous solutions ever better.
Newton converges through LU nonetheless at every grid tried; the practical ceiling in double precision is
$48\times24$ on $|\eta|\le0.9$. The pressure datum $\Pi_0$ is weakly determined by the flux data and is not a
converged observable ($\Pi_0(0.4)-\Pi_0(0)$ still moves at the $10^{-2}$ level between $40\times20$ and
$48\times24$ while $F$ and $U$ converge to $10^{-7}$); the observables are $F$, $U$ and the wall torques.

\begin{table}[h]\centering\small
\begin{tabular}{lcccc}\toprule
grid & $F(0.5,0.4)$ & $U(0.5,0.4)$ & torque at $X_{\rm hi}$ & Newton time \\\midrule
$16\times8$  & 1.142390995 & 0.007421413 & 0.83553529 & 0.3\,s \\
$24\times12$ & 1.141145195 & 0.007428200 & 0.83595640 & 1.0\,s \\
$32\times16$ & 1.141073855 & 0.007428784 & 0.83597306 & 2.4\,s \\
$40\times20$ & 1.141070502 & 0.007428807 & 0.83597372 & 3.7\,s \\
$48\times24$ & 1.141070364 & 0.007428808 & 0.83597374 & 8.9\,s \\\bottomrule
\end{tabular}
\caption{Reference point at $\eta_c=0.9$. Values interpolated at $(X,\eta)=(0.5,0.4)$; the torque is $\partial_X(XF)$ at the outer wall on the dividing plane; the time is a cold Newton solve from rest on an 8-core laptop.}\label{tab:line0}
\end{table}

\section{Parameter sweep}\label{sec:sweep}

\begin{table}[!tp]\centering\small
\setlength{\tabcolsep}{4pt}
\setlength{\arrayrulewidth}{0.3pt}
\begin{tabular}{p{4.0cm}|p{2.6cm}|p{3.8cm}|p{4.6cm}}\toprule
dimensionless group & definition & \GD{} & this work \\\midrule
radial Reynolds number & $\mathrm{Re}_r=|ru_r|/\nu=|V_0|$ & $0$--$20$ (inflow) & $-50.9$ (inflow) to $+50.1$ (outflow) at $\Fhi=1$; to $12$ along the inflow scan at $\Fhi=50$; Line 2 traced at $V_0=-1,-2,-5,-7,-8,-10,-20$ \\
azimuthal Reynolds number & $\mathrm{Re}_\theta=ru_\theta/\nu=2X_{\rm hi}\Fhi\,q^{-A}$ & $0$--$1000$ & $2$--$240$ at the reference instant $q=1$ ($\Fhi=1$--$120$); grows like $q^{-A}$ as the collapse proceeds \\
swirl-to-inflow ratio, outer wall & $u_{to}=V_{to}/|V_{ro}|=2\Fhi q^{-h}/|V_0|$ & $0$--$50$ (example: 50 at $\Rey=1$) & up to $240$ at $V_0=-1$, $8$ at $V_0=-10$ (at $q=1$) \\
radius ratio & $\eta_i=R_i/R_o=\sqrt{X_{\rm lo}/X_{\rm hi}}$ & $0.2$ & $0.02$--$0.51$ (reference $0.2$) \\
anisotropy exponent & $h$, $A=\tfrac12+h$, $D=\tfrac12-h$ & -- (theorem: $0<h<1/100$) & $0.01$--$0.456$ \\
outer similarity radius & $X_{\rm hi}=R_o^2/2q$ & -- & $0.48$--$2.04$ (reference $1$) \\
symmetry-breaking datum & $\delta$ in $F_{\rm hi}(\eta)=\Fhi(1+\delta\eta)$ & $\beta-G$ ($0$ in the example) & $0$--$1$ on the $8\times3$ map (cut range) and on four full-range lines; to $3$ on a few cut-range lines \\
$\eta$ domain & $|\eta|\le\eta_c$ & -- & $\eta_c=0.6,0.8,0.9,1.0$; a cut is admissible only while $|U|<D\eta_c/(1-\eta_c^2)$ \\
grids & $N_X\times N_\eta$ & $N_X+1$ Gauss--Lobatto points in $X$, $N_\eta+1$ in $\eta$ & $16\times8$ to $64\times40$, plain and mapped \\\bottomrule
\end{tabular}
\caption{Dimensionless parameters and the ranges covered. $V_0=ru_r$ is the radial Reynolds number because $\nu=1$; the azimuthal Reynolds number and the swirl-to-inflow ratio are quoted at the reference instant and grow as the collapse proceeds.}\label{tab:params}
\end{table}

\begin{figure}[!tp]\centering
\includegraphics[width=0.9\textwidth]{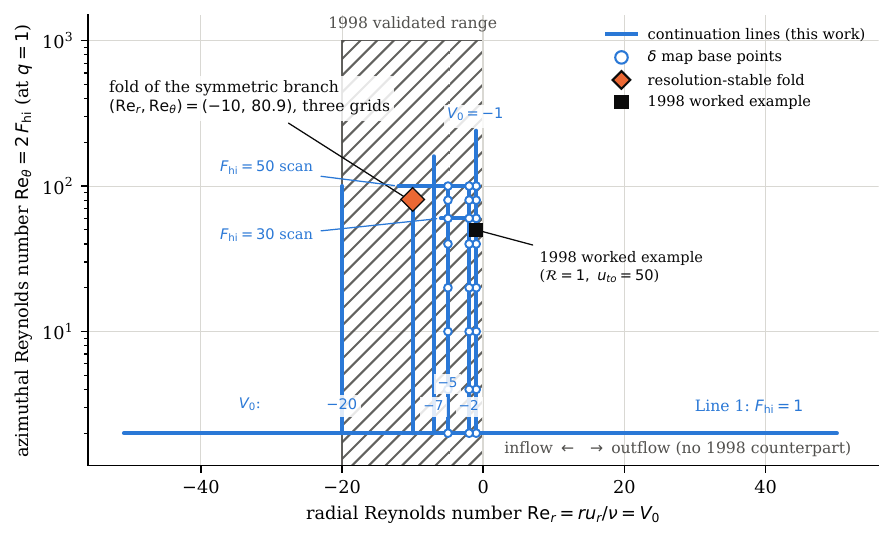}
\caption{The parameter plane covered, in the radial and azimuthal Reynolds numbers at the reference instant.
The hatched rectangle is the range validated in \GD{} (inflow only); the square is its worked example.
Blue segments are the continuation lines of this work and open circles the base points of the $\delta$ map;
the diamond is the fold of the symmetric branch at $(\mathrm{Re}_r,\mathrm{Re}_\theta)=(-10,80.9)$, the one
bifurcation confirmed on three grids and the full $\eta$ range. The outflow half-plane has no \GD{}
counterpart.}\label{fig:params}
\end{figure}

The sweep followed continuation lines from the reference point in the radial Reynolds number $V_0$, the
outer-wall swirl $\Fhi$, the radius ratio $\eta_i$, the anisotropy exponent $h$, the outer similarity radius
$X_{\rm hi}$ and a symmetry-breaking datum $\delta$ in $F_{\rm hi}(\eta)=\Fhi(1+\delta\eta)$; about 140
lines and 6600 converged points were computed on a laptop in one night. We report the results in the
order in which their lessons were learned, because the first lesson governs the rest.

\subsection{The domain rule}\label{sec:rule}
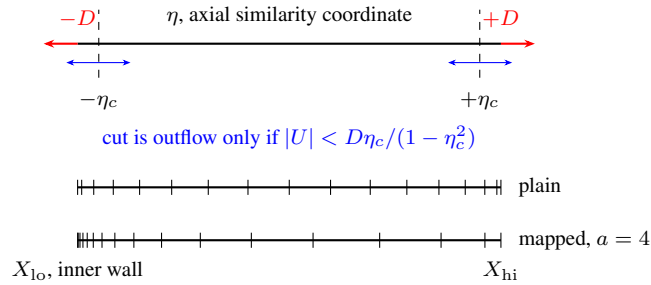
\begin{figure}[!t]\centering
\begin{tikzpicture}[x=1cm,y=1cm]
\draw[thick] (0,3.0)--(5.6,3.0);
\draw[-{Stealth[length=4pt]},red,thick] (0,3.0)--(-0.45,3.0); \draw[-{Stealth[length=4pt]},red,thick] (5.6,3.0)--(6.05,3.0);
\node[anchor=south,red] at (0,3.08) {$-D$}; \node[anchor=south,red] at (5.6,3.08) {$+D$};
\draw[dashed] (0.28,2.55)--(0.28,3.45); \draw[dashed] (5.32,2.55)--(5.32,3.45);
\draw[{Stealth[length=3pt]}-{Stealth[length=3pt]},blue] (4.9,2.75)--(5.75,2.75);
\draw[{Stealth[length=3pt]}-{Stealth[length=3pt]},blue] (-0.15,2.75)--(0.7,2.75);
\node[anchor=north] at (0.28,2.5) {$-\eta_c$}; \node[anchor=north] at (5.32,2.5) {$+\eta_c$};
\node[anchor=south] at (2.8,3.1) {$\eta$, axial similarity coordinate};
\node[anchor=north,blue] at (2.8,2.05) {cut is outflow only if $|U|<D\eta_c/(1-\eta_c^2)$};
\draw[thick] (0,1.1)--(5.6,1.1);
\foreach \i in {0,...,16}{\pgfmathsetmacro{\xx}{5.6*(1-cos(180*\i/16))/2}\draw (\xx,1.0)--(\xx,1.2);}
\node[anchor=west] at (5.7,1.1) {plain};
\draw[thick] (0,0.4)--(5.6,0.4);
\foreach \i in {0,...,16}{\pgfmathsetmacro{\uu}{(1-cos(180*\i/16))/2}\pgfmathsetmacro{\xx}{5.6*sinh(4*\uu)/sinh(4)}\draw (\xx,0.3)--(\xx,0.5);}
\node[anchor=west] at (5.7,0.4) {mapped, $a=4$};
\node[anchor=north] at (0,0.25) {$X_{\rm lo}$, inner wall}; \node[anchor=north] at (5.6,0.25) {$X_{\rm hi}$};
\end{tikzpicture}
\caption{Top: the axial similarity coordinate with the cut at $\pm\eta_c$; at the singular ends the transport coefficient is $\pm D$ and the flow is outward for any axial profile, at the cut it is outward only while $|U|<D\eta_c/(1-\eta_c^2)$. Bottom: sixteen radial nodes on the plain Chebyshev grid and on the sinh-mapped grid clustered at the inner wall, where the strong-inflow layer sits.}\label{fig:gridsweep}
\end{figure}
A cut of the axial range is admissible only while the axial profile is small at the cut (Figure~\ref{fig:gridsweep}). The independence of the solution from the cut $|\eta|\le\eta_c$ observed at the reference point holds only
while the cut is an outflow boundary, i.e.\ while $D\eta+dU$ points outward at $\eta=\pm\eta_c$ for every
$X$. This requires
\begin{equation}\label{eq:rule}
|U|<\frac{D\eta_c}{1-\eta_c^2}\qquad\text{at }\eta=\pm\eta_c,
\end{equation}
which is $2.32$ for $\eta_c=0.9$. At the reference point $\max|U|=0.06$. Along the swirl line at $V_0=-1$
the condition fails from $\Fhi\approx8$ on: at 7 of 33 radial nodes for $\Fhi=10$ ($\max|U|=4.0$), 13 of
33 at $\Fhi=20$, 15 of 33 at $\Fhi=50$ ($\max|U|=22$). Characteristics then enter through the cut, the
truncated problem lacks boundary data, and the collocation supplies them implicitly. The effect is
measurable and systematic: at $(\Fhi,V_0)=(50,-1)$ the interpolated $F(0.5,0.4)$ converges to $10.614$ on
$|\eta|\le0.9$ and to $10.925$ on the full range ($32\times16$, $40\times20$, $48\times24$ each), $3\%$
apart. On the truncated domain the sweep produced an elaborate bifurcation structure: an S-shaped symmetric
branch in swirl with folds at $\Fhi=22.7$ and $66.95$ for $V_0=-1$ and $21.0$ and $68.3$ for $V_0=-2$,
three coexisting symmetric solutions between the folds, symmetry-breaking pitchforks on the middle sheets
and an asymmetric branch bridging two of them, and snaking in narrow parameter windows above $\Fhi=50$.
None of it exists on the full range. At $\eta=\pm1$ the transport coefficient is $\pm D$ for any $U$, since
$d=0$, so no data are needed there, and the full range is the correct domain whenever $|U|$ can exceed
about $2$; the price is algebraic rather than geometric convergence in $\eta$, about $1\%$ at $32\times16$
for $\Fhi=50$. We record \eqref{eq:rule} as a result of method and now log the outflow margin with every
solve.

\subsection{Results on the full range}\label{sec:full}
With $\eta_c=1$, $32\times16$ as the working grid and $40\times20$ and $48\times24$ as checks, the
symmetric self-similar swirl between contracting porous walls is a single smooth branch in every direction
swept at inflow below about $9$; the structure at stronger inflow is described in
Section~\ref{sec:layer}.
\begin{itemize}[leftmargin=*]
\item Radial Reynolds number: fold-free from $V_0=-50.9$ (inflow) to $+50.1$ (outflow) at $\Fhi=1$, on
$32\times16$, $40\times20$ and $48\times24$.
\item Outer swirl: fold-free from $\Fhi=1$ to $120$ at $V_0=-1$, 235 continuation points with step
$\le0.3$, identical on $32\times16$ and $40\times20$; fold-free to $\Fhi=50$ at $V_0=-2$ and $-5$
($48\times24$ at $-5$).
\item Inflow at fixed swirl: fold-free from $V_0=-1$ to $-6$ at $\Fhi=30$ and to $-9$ at $\Fhi=50$
($48\times24$).
\item Radius ratio $\eta_i$ from $0.51$ to $0.02$ ($40\times20$), $h$ from $0.01$ to $0.456$, $X_{\rm hi}$
from $0.48$ to $2.04$: fold-free.
\item The reflection symmetry $U(X,0)=0$ holds to $10^{-13}$ on every point of every symmetric line.
\end{itemize}
One candidate bifurcation survived the change of domain: a determinant sign change at $\Fhi=50$ with the
parity signature of a pitchfork (null vector with $F$ odd and $U$ even; projection of $\partial_{V_0}R$ on
the left null vector $6\cdot10^{-16}$) and an asymmetric solution at $\delta=0$ on one side. It sits at
$V_0=-5.75$ on $32\times16$, at $-7.8$ on $40\times20$, and is absent down to $V_0=-9$ on $48\times24$. A
bifurcation that recedes with every refinement is an artifact, here of the wall layer of
Section~\ref{sec:layer}, and we record it as such.

\paragraph{Response to asymmetry.} The response of the dividing-plane axial velocity to the symmetry-breaking datum is single-valued and finite over the whole map, and its linear part changes sign near a swirl of $10$ to $30$. The datum $\delta$ breaks the reflection symmetry as $\beta-G$ does in
\eqref{eq:98sym}. On the full range all 24 continuation lines in $\delta$ from $0$ to $1$ (eight swirls,
three inflows) are single-valued; a singular event seen at $32\times16$ near $(\Fhi,V_0,\delta)=(40,-5,0.75)$
is absent at $40\times20$. Table~\ref{tab:slopes} gives the linear response of the dividing-plane axial
velocity, $\partial U(X_{\rm mid},0)/\partial\delta$ at $\delta=0$. It grows roughly linearly with swirl to
a peak near $\Fhi=10$, changes sign between $\Fhi=10$ and $20$ for $V_0=-2,-5$ and between $20$ and $30$ for
$V_0=-1$, and then grows in magnitude with the opposite sign, faster at stronger inflow. At $V_0=-1$ and $-2$
the truncated-domain sweep had given the same numbers to within a few percent, so the linear response is
robust to the cut even where the interior is not; at $V_0=-5$ and $\Fhi\ge20$ it was not.

\begin{table}[h]\centering\small
\begin{tabular}{lcccccccc}\toprule
$V_0$ & $\Fhi=1$ & 2 & 5 & 10 & 20 & 30 & 40 & 50 \\\midrule
$-1$ & 0.111 & 0.424 & 1.94 & 2.94 & 0.709 & $-0.945$ & $-1.87$ & $-2.40$ \\
$-2$ & 0.161 & 0.618 & 2.77 & 3.18 & $-0.820$ & $-2.80$ & $-3.94$ & $-4.73$ \\
$-5$ & 0.183 & 0.729 & 4.43 & 13.8 & $-6.53$ & $-7.01$ & $-9.32$ & $-16.9$ \\\bottomrule
\end{tabular}
\caption{Linear response $\partial U(X_{\rm mid},0)/\partial\delta$ at $\delta=0$, full $\eta$ range,
$32\times16$ (first two points of each continuation line; $X_{\rm mid}=0.52$). The $V_0=-5$ row is in the
regime of Section~\ref{sec:layer} and carries its uncertainty.}\label{tab:slopes}
\end{table}

\subsection{A wall layer at strong inflow}\label{sec:layer}
At strong wall inflow the layer that limits resolution is radial and sits at the inner wall. At $|V_0|\gtrsim8$ the outer-wall torque $\partial_X(XF)$ does not converge on any unmapped grid we
tried. At $V_0=-20$, $\Fhi=1$, with $N_\eta=24$ it is $-0.60$, $-0.081$, $-0.0093$, $-0.0005$ and $+0.0004$
for $N_X=32,40,48,56,64$, while refining $\eta$ from 16 to 40 modes at $N_X=48$ leaves $F$, $U$ and the
torque unchanged to $10^{-6}$ or better: the layer is radial. Its location follows from a sinh-mapped
$X$ grid, $X(s)=X_{\rm lo}+(X_{\rm hi}-X_{\rm lo})\,\sinh(au)/\sinh(a)$ with $u=(1+s)/2$, which clusters
nodes at the inner wall for $a>0$ (and, mirrored, at the outer wall). Clustering at the outer wall makes
the torque worse at every $N_X$ ($-20$, $-6.6$, $-2.0$ for $N_X=24,32,40$ at $a=2$); clustering at the inner
wall with $a=4$ gives $+3.95\cdot10^{-4}$, $+5.23\cdot10^{-4}$, $+5.24\cdot10^{-4}$ at $N_X=24,32,40$, with the
inner-wall torque and $\max|U|$ converged to seven and five digits. The layer sits at the inner porous
wall, where the fluid leaves the annulus under inflow, and 32 mapped modes resolve at $V_0=-20$ what 64
plain modes did not. The same map reproduces the reference-point torque of Table~\ref{tab:line0} to eight
digits at $32\times16$: the slow inner-wall convergence seen there is the same layer in mild form. This is
why the candidate of Section~\ref{sec:full} moved with resolution on plain grids. On the mapped grid and the
full range the swirl line at $V_0=-10$ is no longer fold-free: the symmetric branch ends in a fold at
$\Fhi=40.466$ ($32\times16$) and $40.467$ ($40\times20$), with the dividing-plane slope $U_\eta(X_{\rm mid},0)$
($9.79$ and $9.70$) and the outer torque ($0.111$ at both) matching at the fold. This is the one bifurcation
of the study that survives both the domain rule and a resolution check. At $48\times24$ the fold is at $\Fhi=40.468$ with the same torque, so its location is converged to
$5\cdot10^{-5}$. The branch that returns from the fold, where $\max|U|$ reaches $60$, differs between all
three grids (slope $17.5$, $25.7$, $10.9$ at $\Fhi=45$); refining $X$ alone from 48 to 64 modes leaves it
unchanged to four digits, refining $\eta$ alone from 24 to 40 modes moves the torques by 5--8\%, and refining to 56 modes leaves the torques oscillating at the few-percent level, so on
that sheet the solution is not converged in $\eta$ at any resolution tried; with $\max|U|\sim50$ the
coefficient $D\eta+dU$ of $F_\eta$ changes sign inside the domain, which moves the analytic-class
non-uniqueness of the $\eta$ direction from the ends into the interior and is the likely cause; at $V_0=-5$ the
swirl line is fold-free to $\Fhi=50$ on the same grid, and at $-20$ to $50$, where the fold law below puts the
fold at $164$ (it is found at $170.6$ on the fine grid).

\paragraph{Fold locus.} The fold lies on $\Fhi^*\simeq0.41\,V_0^2$, and its location is a converged object: on the mapped full-range grid the
$48\times24$ and $56\times28$ folds coincide with the $32\times16$ ones to four digits at every inflow tried.
Tracing the swirl line at fixed wall inflow gives the upper fold at $\Fhi^*=33.01$, $40.47$, $49.07$, $58.71$,
$93.49$ and $170.59$ for $V_0=-9,-10,-11,-12,-15,-20$ (Table~\ref{tab:fold}); the first five are the same on the
coarse and the fine grids, the last was reached only on $48\times24$, where the base point at that inflow
needs a homotopy in swirl and flux together. The ratio $\Fhi^*/V_0^2$ drifts slowly upward, from $0.405$ at
$|V_0|=10$ to $0.427$ at $20$, so the fold lies on
\begin{equation}
\Fhi^*\simeq0.41\,V_0^2,\qquad\text{i.e.}\qquad \mathrm{Re}_\theta^*\simeq0.82\,\mathrm{Re}_r^2
\label{eq:foldlaw}
\end{equation}
to within $4\%$ for $9\le|V_0|\le20$, and to $0.35\%$ on $\Fhi^*=0.488V_0^2-1.62|V_0|+8.0$ or, equivalently,
$\Fhi^*/V_0^2=0.386+0.0020|V_0|$. At $V_0=-7$ and $-8$ the line is fold-free to $\Fhi=80$ (to $60$ at
$40\times20$); the S-curve is born near $V_0\approx-8.7$, where at $-9$ it is only $0.7$ wide ($33.01$ and
$32.3$). In words: once the radial Reynolds number exceeds about $9$, the steady self-similar swirl between
porous walls ceases to exist beyond a critical swirl that grows as the square of the inflow, and the wall
torques on the symmetric sheet agree between grids to $0.1\%$ up to the fold.

\begin{table}[h]\centering\small
\begin{tabular}{rrrrrl}\toprule
$|V_0|$ & $\Fhi^*$ ($32\times16$) & $\Fhi^*$ (fine grid) & $\Fhi^*/V_0^2$ & $\mathrm{Re}_\theta^*=2\Fhi^*$ & lower fold \\\midrule
9  & 33.01 & 33.01 ($48\times24$) & 0.4075 & 66  & 32.36; 32.29 (fine) \\
10 & 40.47 & 40.47 ($56\times28$) & 0.4047 & 81  & 37.33; 37.42 (fine) \\
11 & 49.06 & 49.07 ($48\times24$) & 0.4055 & 98  & 42.75 (fine only) \\
12 & 58.71 & 58.71 ($48\times24$) & 0.4077 & 117 & 45.12; 48.64 (fine) \\
15 & 93.48 & 93.49 ($48\times24$) & 0.4155 & 187 & not captured \\
20 & --    & 170.59 ($48\times24$) & 0.4265 & 341 & not captured \\\bottomrule
\end{tabular}
\caption{The fold of the symmetric branch on the mapped full-range grid ($X$ map parameter $-4$, $\eta_c=1$):
critical swirl at fixed radial Reynolds number $|V_0|$ on the working and the fine grids, the ratio to $V_0^2$,
the azimuthal Reynolds number at the reference instant, and the lower fold that closes the hysteresis window
where it was reached. The upper fold is grid-converged; the lower fold and the returning sheet are not.}
\label{tab:fold}
\end{table}

\paragraph{Beyond the fold.} What replaces the steady state beyond the fold is not resolved by this
discretization. On every grid the continuation turns at the fold onto a returning sheet, and on the finer grids
that sheet is a different object: between $32\times16$ and $48\times24$ the outer-wall torque on it differs by
$27$--$55\%$ and the inner-wall torque by $2$--$5\%$ at $|V_0|=9$ to $12$, against $0.1\%$ on the symmetric sheet;
the lower fold that closes the hysteresis window moves ($45.1\to48.6$ at $|V_0|=12$); at $|V_0|=15$ the
$48\times24$ continuation meets a singular Jacobian at the fold (smallest bordered singular value $10^{-15}$)
and retraces the symmetric branch downward, reproducing its torques to $10^{-6}$; and at $|V_0|=20$ the
returning sheet, with axial velocities $50\%$ larger than on the symmetric one, stalls at $\Fhi=97.8$ on a
singular Jacobian after 99 points. The axial velocity on these sheets reaches $50$--$330$ in similarity units,
so the transport coefficient $D\eta+dU$ changes sign well inside the domain and the collocation is no longer
selecting the analytic branch (Section~\ref{sec:sweep}). The existence and location of the fold are therefore
results; the state beyond it is not, and a formulation that imposes the analytic branch across interior sign
changes is the prerequisite for computing it.

\subsection{Stability of the self-similar swirl}\label{sec:stab}
\begin{figure}[!t]\centering
\includegraphics[width=\textwidth]{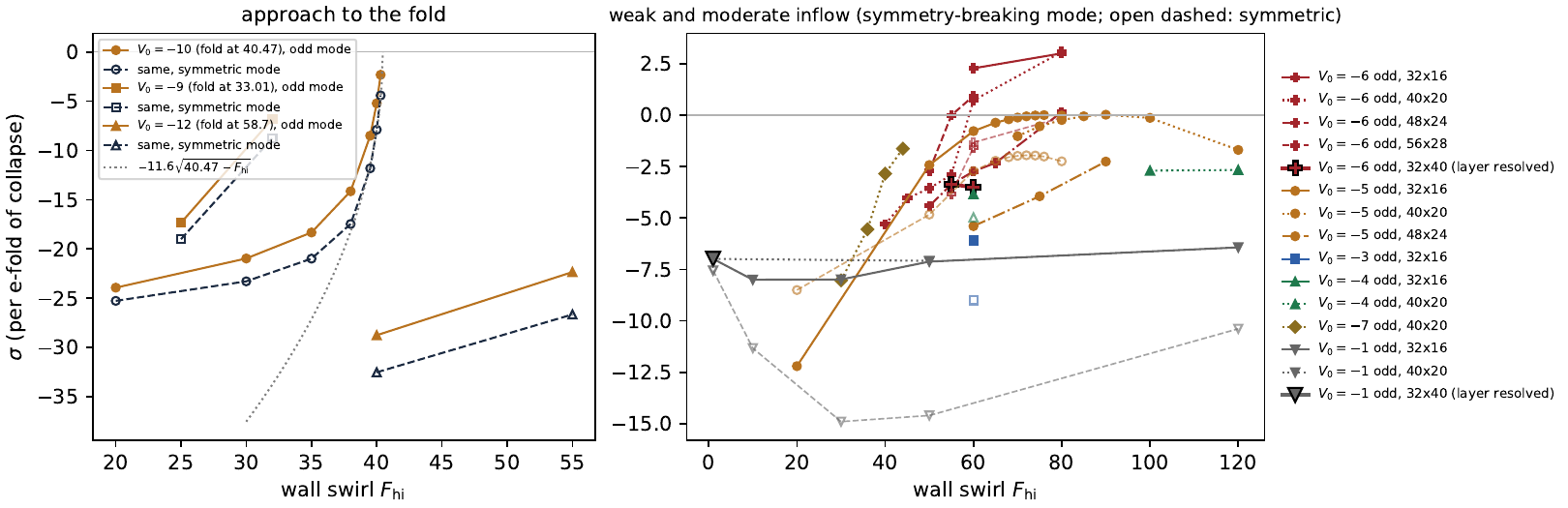}
\caption{Leading eigenvalues $\sigma$ of the self-similar swirl between porous walls, in e-folds of collapse (negative: the perturbation decays). Left: the approach to the fold along lines of fixed inflow; the symmetric mode (dashed) follows the saddle-node law $-11.6\sqrt{F^*_{\rm hi}-F_{\rm hi}}$ at $V_0=-10$, the symmetry-breaking mode (solid) runs ahead of it. Right: weak and moderate inflow; along $V_0=-6$ the least-damped eigenvalue at $F_{\rm hi}\ge55$ scatters with the number of $\eta$ nodes ($32\times16$ solid, $40\times20$ dotted, $48\times24$ dash-dotted, $56\times28$ dashed, $32\times40$ and $32\times56$ large markers) and does not converge (text: the sonic layer at $|\eta|\approx0.99$); at $V_0\le-5$ and $F_{\rm hi}\le50$ the grids agree.}\label{fig:stab}
\end{figure}

Whether the profile attracts is a question the steady problem cannot answer, and it decides whether a real collapse selects it. We add the collapse time to the leading-order system by dynamic rescaling: with $s=-\ln q$, one unit of $s$ being one e-fold of collapse, $\partial_t(q^bF)=q^{b-1}(T_bF+L^{-1}F_s)$, so the equations become $L^{-1}F_s+R_\theta=0$ and $L^{-1}U_s+R_z=0$ with continuity, the radial balance, the wall data and the gauge as algebraic constraints, and the steady states are exactly the profiles of the sweep. We march this system with a second-order backward-difference method, one complex-step Jacobian per step, and we solve the generalized eigenproblem $\sigma Mv=-Jv$ for the spectrum, $M$ the mass matrix of the two evolution equations and $J$ the Jacobian we already form; a fold is a zero eigenvalue of this problem, and the march's decay rate along an eigenvector reproduces its eigenvalue to $0.2\%$ (tests in \texttt{tests\_stability.py}). Figure~\ref{fig:stab} collects the leading eigenvalues. The self-similar swirl between porous walls is an attractor of the collapse at weak inflow and along the approach to the fold: at the reference point the leading eigenvalue is $\sigma_1=-6.987$ per e-fold, the same on $20\times10$, $32\times16$ and $40\times20$ to four digits, the whole leading spectrum is real and negative, and asymmetric wall data ($\delta$ up to $0.5$) or strong inflow ($\sigma_1=-84$ at $V_0=-20$) do not change the verdict; a perturbation dies by a factor $e$ in a seventh of an e-fold of collapse. The least-damped mode is almost always the symmetry-breaking one, odd in the swirl and even in the axial velocity. Along the swirl line at $V_0=-10$ the spectrum sees the fold independently of the continuation: the symmetric mode's eigenvalue follows $\sigma_{\rm even}\simeq-11.6\sqrt{F^*_{\rm hi}-F_{\rm hi}}$ with $F^*_{\rm hi}=40.47$ ($-11.8$, $-7.9$, $-4.4$ at $39.5$, $40.0$, $40.3$), the signature of a saddle-node, and the symmetry-breaking mode runs ahead of it ($-8.5$, $-5.2$, $-2.3$ at the same points), extrapolating to zero just past the fold; the same ordering holds at $V_0=-9$. So the sheet that returns from the fold should break the reflection symmetry almost at once, which is what the determinant sign changes within $0.1$ of the fold recorded in Section~\ref{sec:layer} and the non-convergence of that sheet under a symmetric collocation both suggest. At moderate inflow and strong swirl the spectrum is harder to converge, and the reason is a second result of method. In the inner half of the annulus the axial velocity next to the outflow boundaries is directed toward the mid-plane ($U\approx-30$ at $\eta\to1$ for $V_0=-6$, $F_{\rm hi}=55$), so the coefficient of $F_\eta$ in the swirl equation, $c=D\eta+U(1-\eta^2)$, is negative over most of the half-domain and turns positive only in a layer $1-\eta^2\approx D/|U|$, about $0.016$ wide, against the boundary: the characteristics diverge from the sonic line $c=0$, which is a regular singular point of the first-order $\eta$ operator. The steady profile converges only algebraically there, and the linearized operator carries modes concentrated in the layer whose eigenvalues depend on how many collocation nodes the layer contains. Chebyshev's first interior node sits at $1-\cos(\pi/N_\eta)$, so $k$ nodes fall inside the layer only for $N_\eta>\pi k\sqrt{|U_b|/D}$, about $25k$ for $|U_b|=30$; with $16$ to $28$ nodes the layer holds none, and at $(V_0,F_{\rm hi})=(-6,60)$ the least-damped symmetry-breaking eigenvalue reads $+2.28$, $+0.72$, $-2.7\pm4.6i$ and $+0.94$ on $16$, $20$, $24$ and $28$ nodes, each grid's nonlinear march agreeing with its own spectrum (the $20$-node grid even settles on an asymmetric steady profile with a net axial flow through the dividing plane), while the number of nodes in $X$ changes nothing. Refining further does not settle it: with $32$, $40$, $48$, $56$ and $64$ nodes the least-damped eigenvalue at $(-6,60)$ is $+3.1$, $-3.5$, $-1.8$, $+7.7$ and $+5.0$ (the last two a symmetric mode), and a $\tanh$ map of the $\eta$ nodes that places five or more nodes in the layer, and which reproduces the reference spectrum to four digits, gives $+3.9$ and $+2.3$ with $28$ and $40$ nodes at one map strength and $-0.3$ and $+0.2$ at another, against $-3.1$ on the plain grid with $80$ nodes at $(-6,55)$, while the steady profile itself converges (its maximum axial velocity changes by $2\%$ over the same sequence). Every one of these wandering modes peaks at the inner-wall side of the sonic line, $X\approx0.08$, $|\eta|\approx0.99$, and the ones on the finest plain grids oscillate from node to node there. The steady profile is affected too, though less: at $(-6,40)$ its maximum axial velocity varies by $5\%$ between $16$ and $32$ nodes on plain and mapped grids, and its leading eigenvalue between $-6.3$ and $+1.9$, whereas at $(-2,10)$, where the sonic line lies at $|\eta|\approx0.92$ and $|U|\approx6$, profile and spectrum agree to four digits from $8$ nodes on; so the difficulty sets in between axial velocities of $6$ and $34$ at the sonic line and the sweep's values at moderate inflow and strong swirl carry an uncertainty of a few percent. At $(-6,55)$ and $(-6.5,55)$ the same grids give damped spectra from $40$ nodes on, but the scatter at $60$ warns against reading that as convergence. Our reading is that the linearized operator at moderate inflow and strong swirl carries a continuous spectrum generated at the sonic line, where power-law modes $|\eta-\eta_*|^{\gamma(\sigma)}$ are admissible for a half-plane of $\sigma$, and that each grid samples it at its own places; whether that half-plane reaches into $\mathrm{Re}\,\sigma>0$, which would make the sonic line the seat of a non-modal instability of the collapse at these parameters, is the open question; the sequences at lower swirl and at weak inflow reported above locate its onset. The statements that are grid-converged are therefore these: the profile is an attractor at weak inflow and along the whole approach to the fold, where profiles and eigenvalues agree on $16$ to $32$ nodes although a sonic line is present there too (at $(-10,40)$ the axial velocity at $\eta\to1$ reaches $-67$ near the inner wall), so the presence of the line is necessary but not sufficient for the scatter and the distinction between the two regimes is not yet understood; the fold is a saddle-node; and at moderate inflow with strong swirl the spectral question is open, the earlier reading of a symmetry-breaking instability there being one grid's sample of the scatter.

\section{The axis problem}\label{sec:axis}
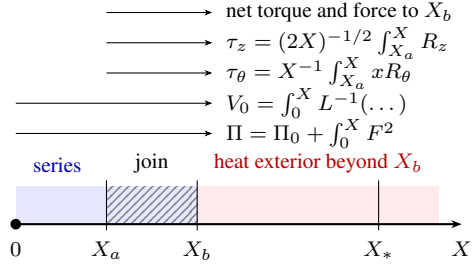
\begin{figure}[!t]\centering
\begin{tikzpicture}[x=1cm,y=1cm]
\fill[blue!10] (0,0) rectangle (4.8,0.5);
\fill[pattern=north east lines,pattern color=gray] (1.2,0) rectangle (2.4,0.5);
\fill[red!8] (2.4,0.5) rectangle (5.6,0.0);
\fill[pattern=north east lines,pattern color=gray] (1.2,0) rectangle (2.4,0.5);
\draw[thick,-{Stealth[length=4pt]}] (0,0)--(5.9,0);
\fill (0,0) circle (2pt);
\foreach \x/\l in {1.2/$X_a$,2.4/$X_b$,4.8/$X_*$}{\draw (\x,-0.08)--(\x,0.58); \node[anchor=north] at (\x,-0.1) {\l};}
\node[anchor=north] at (0,-0.1) {$0$}; \node[anchor=north] at (5.9,-0.1) {$X$};
\node[anchor=south,blue!70!black] at (0.55,0.55) {series};
\node[anchor=south] at (1.8,0.55) {join};
\node[anchor=south,red!70!black] at (4.0,0.55) {heat exterior beyond $X_b$};
\foreach \y/\x/\l in {1.2/0/{$\Pi=\Pi_0+\int_0^X F^2$},1.6/0/{$V_0=\int_0^X L^{-1}(\dots)$},2.0/1.2/{$\tau_\theta=X^{-1}\int_{X_a}^X xR_\theta$},2.4/1.2/{$\tau_z=(2X)^{-1/2}\int_{X_a}^X R_z$},2.8/1.2/{net torque and force to $X_b$}}{
\draw[-{Stealth[length=3pt]}] (\x,\y)--(2.6,\y); \node[anchor=west] at (2.65,\y) {\l};}
\end{tikzpicture}
\caption{The axis problem as a Cauchy problem in the viscous similarity radius: the series converges on $0\le X<X_*$ (blue), the join to the heat exterior (red) is made across the annulus $X_a<X<X_b$ (hatched), and the five cumulative integrals that close the system, the similarity pressure and the wall inflow from $X=0$ and the two pulse stresses and their net torque and force from $X_a$, run outward.}\label{fig:gridaxis}
\end{figure}
The Dirichlet form of the axis problem has no resolution-stable solution; posed as a Cauchy problem in $X$ from analytic axis data, as the \OAI{} construction itself does, it has one, and its series, march and join stages are implemented and verified (Figure~\ref{fig:gridaxis}). The core of the \OAI{} construction lives on the axis domain $X\in[0,X_{\rm hi}]$. Posed as a Dirichlet problem
($F$ and $U$ given at $X_{\rm hi}$, $\Pi_0$ given, regularity at $X=0$ imposed by collocating the
degenerate equations there), Newton converges quadratically at fixed resolution but to different solutions
on different grids. Deflated Newton \cite{fbf15}, which divides the residual by the distance to every
solution already found and so cannot return to one, makes the situation precise: with $\Fhi=1$,
$X_{\rm hi}=2$ and six initial guesses per grid it finds, on $|\eta|\le0.8$, one solution at $12\times6$
($F(0,0)=0.963$), none at $16\times8$, four at $20\times10$ ($F(0,0)=0.871$, $0.670$ and two with swirl
overshooting to $2.5$), and one at $24\times12$ ($0.519$); on the full range none, none, two ($0.537$ and
$-792$) and none. No solution persists across grids at either cutoff. The pre-2026 solver reproduces the
three numbers first reported for this problem exactly, so the defect is in the formulation. The \OAI{}
construction itself (its Proposition B.2) does not pose the axis problem this way: it prescribes analytic data on
the axis and continues outward. Expanding $F=\sum_kF_k(\eta)X^k$, $U=\sum_kU_k(\eta)X^k$, with $v_k$ and
$P_k$ following algebraically from continuity and radial balance, the viscous terms are the only ones that
raise the power of $X$, and \eqref{eq:core} becomes an explicit recursion in $k$,
\begin{align}
2(k+1)(k+2)\,F_{k+1}&=L^{-1}\!\big(-(A+\tfrac12)F_k+D\eta F_k'+kF_k\big)+\mathcal N^{(k)}_F[F,U,v],\label{eq:R1}\\
2(k+1)^2\,U_{k+1}&=L^{-1}\!\big(-AU_k+D\eta U_k'+kU_k\big)+\big[Z_{-2A}\Pi\big]_k+\mathcal N^{(k)}_U[F,U,v],\label{eq:R2}
\end{align}
where $\mathcal N^{(k)}$ collects the products of lower-order coefficients at order $X^k$ and
$[\,\cdot\,]_k$ the order-$X^k$ part. The free data are $F_0(\eta)$, $U_0(\eta)$ and $\Pi_0(\eta)$. The
coefficient growth gives the radius of convergence in $X$, beyond which the series is continued by an
ordinary-differential-equation march in $X$ with $\eta$ collocated; the sideways problem amplifies the
$\eta$-mode of wavenumber $k$ like $\exp(c\sqrt kX)$, a few e-folds for the resolutions in use, and the
low-dimensional parametrization of the axis data keeps the high modes unexcited. The series stage is
implemented and verified (\texttt{axis\_series.py}): with the forcing of a manufactured solution the
recursion reproduces its Taylor coefficients to $10^{-14}$--$10^{-12}$ for $k\le10$ on 16 to 32 $\eta$ modes;
even $F_0$, odd $U_0$ and even $\Pi_0$ give $F_k$ even and $U_k$ odd to $10^{-12}$ for $k\le20$; for the axis
data $F_0=1$, $U_0=\tfrac12\eta$, $\Pi_0=0$ the coefficients decay geometrically with a radius of convergence
$X_*=3.9$--$4.0$ (ratio and root tests, independent of the $\eta$ resolution), and the partial sum to $k=40$
satisfies the full leading-order residuals to $10^{-10}$ on $X\in[0,1.17]$, so for such data the series alone
reaches the join radius and the march is needed only when $X_*$ is smaller. One fact about the exterior came
out of the same test: $E=c_\infty X^{-A}H(2d/X)$ is not smooth at $X=0$ in the profile variables, since
$H(Z)\sim Z^{-h}/\Gamma(1+h)$ as $Z\to\infty$ gives $F\sim0.706/X$, the potential vortex; core and exterior
therefore meet in the annulus and are not analytic continuations of each other. The join is implemented as well (\texttt{axis\_join.py}): the series core is blended into the heat exterior
across $1<X<2$ with a $C^\infty$ partition of unity, the exterior amplitude $c_\infty$ is fitted to the core swirl
at the outer edge, and the exterior pressure $\Pi_{\rm ext}=-\int_X^\infty F_{\rm ext}^2$ fixes the axis pressure
datum, so the free axis data are $(F_0,U_0,c_\infty)$. The residuals of the blended profile vanish to $10^{-10}$
in the core and $10^{-6}$ in the exterior and are $O(1)$--$O(10)$ in the annulus; integrated radially they give the
pulse stresses $\tau_\theta=X^{-1}\int_{X_a}^X xR_\theta\,dx$ and $\tau_z=(2X)^{-1/2}\int_{X_a}^X R_z\,dx$, with peaks
$0.53$ and $1.18$ for $F_0=0.1$, $U_0=0.05\,\eta$. The stress coefficients of \OAI{}, its cone condition and its matching functional are implemented
and verified as well (\texttt{axis\_cone.py}, 27 checks). \OAI{} writes the stress the pulses must supply as
$T_0=F(p_s-s)$ with $s=(a,-b_s)$, $a=-2XF_X/F$ the logarithmic swirl gradient and $b_s=2XU_X/E$ the shear ratio,
and $p_s$ built from five cumulative radial integrals of the profile (its eqs.~(4.11), (4.15)--(4.16)); on the
annulus this reproduces the residual-integrated stresses above to $3\cdot10^{-7}$ relative to their peaks, on the
series core it vanishes to $10^{-8}$, and on the pure heat exterior it reduces to the boundary terms at $X_b$ to
$10^{-6}$. One physical fact came out of that check: the exterior with $U=0$ is stress-free only when the axial
moment $M=\int_0^X U\,dx$ vanishes at $X_b$, since otherwise the carried flux transports swirl; after matching,
$M(\infty)=0$ removes the term.

\paragraph{The march, and two rules of resolution.} The Cauchy problem in $X$ is ill posed in the sense of
Hadamard, with a measured growth law, and the loss of the stress identity at large amplitude was an end layer,
not a resolution limit. The march is implemented (\texttt{axis\_march.py}): from $X_s=0.4X_*$ the state, consisting of $F$, $F_X$, $U$, $U_X$, $\Pi$ and $Xv_0$ on
the $\eta$ nodes, is integrated in $X$ by an eighth-order Dormand--Prince method with
tolerance $10^{-11}$; it agrees with the series at $0.7X_*$ to the series' own truncation ($10^{-10}$ to $10^{-8}$),
preserves parity to $10^{-11}$ and satisfies the leading-order residuals to $6\cdot10^{-8}$ on a 64-node Chebyshev
grid. A perturbation in Chebyshev mode $m$ of $\eta$ grows like $\exp\sqrt{8cmX}$ with $c=0.36$ measured, so roundoff
in mode 24 is amplified $2\cdot10^6$ by $X=3$, the unfiltered recursion diverges for $N_\eta\ge32$, and every stage
projects onto the first 28 modes at most; the pressure fixed point that appeared to break at $N_\eta=32$ was this
divergence, and Anderson mixing converges it in 7 to 18 iterations otherwise. The domain rule of
Section~\ref{sec:sweep} reappears at the singular ends: axis data with axial inflow toward the dividing plane
beyond about $0.3$, with swirl reversing sign in $\eta$, or with $U_0$ vanishing at $\eta=\pm1$ steepen at the
ends without limit ($U_\eta(1,\pm1)=74$ for the step-3 optimum, no two filters agreeing there), while axial outflow
$U_0=a\eta$ is regular to $a\ge1$ with the identity satisfied to $10^{-8}$; the \OAI{} core, spun up by the
inflow and evacuated along the axis, is of the regular kind.

\paragraph{Cone and matching.} The cone condition fails on the whole annulus for every smooth blend we computed,
and the smooth blend admits no nontrivial matched axis data; both are expected, and both bound what this core
representation can reach. With $t_s=-b_s/a$, $v_s=a(1+t_s^2)$, $P_c=p_{s,1}+t_sp_{s,2}$ and
$J_c=p_{s,2}-t_sp_{s,1}$, the admissible cone of eqs.~(4.20)--(4.22) of \OAI{} is $P_c>v_s$ together with
$(v_s-2)J_c^2<2(P_c-v_s)^2$, and its eq.~(4.26) asks for a uniform margin $\kappa\in(0,2)$ on the closed annulus,
\begin{equation}\label{eq:cone}
n_\theta+t_sn_z\ge\kappa,\qquad (v_s-2)(n_z-t_sn_\theta)^2\le(2-\kappa)(n_\theta+t_sn_z)^2,
\end{equation}
for the unit stress direction $n=T_0/|T_0|$, with $n\parallel(a,-b_s)$ at $X_a$ and $n=(1,0)$, $b_s=0$ at $X_b$.
For the sample data $a>0$ only on the outer $40\%$ of the annulus ($X\ge1.66$, where the swirl turns over toward
the exterior), $P_c<0$ throughout, no point is admissible, $\kappa=-74$, and $n=(-1,0)$ at $X_b$: the net torque
of the smooth blend has the wrong sign. The matching functional of Theorem~4.6(v) of \OAI, six functions of
$\eta$ (net torque and net force at $X_b$, $M(\infty)=J(\infty)=S(\infty)=0$ and $\int_0^\infty(H-H_{\rm pow})\,dX=0$),
was minimized by least squares over $F_0\in{\rm span}(T_0,T_2,T_4)$ and $U_0\in{\rm span}(T_1,T_3)$. With $c_\infty$
free the fit collapses onto the trivial profile ($c_\infty\to10^{-7}$, all residuals below $5\cdot10^{-7}$): the
identities do not select an amplitude, and $c_\infty$ is a normalization that must be pinned. With $c_\infty=0.2$ the
fit stalls at a root-mean-square residual of $0.44$, dominated by $S(\infty)=\int_0^\infty(U^2-E^2/2)\,dX$, whose
exterior tail $c_\infty^2X_b^{-2h}/(4h)$ is $0.99$, and the marched core at $N_\eta$ up to 40 stalls at the same
optimum to three digits, widened to seven axis parameters or given free annulus content alike. The obstruction is
exact, and it is a statement about the flow. Theorem~4.6 of \OAI{} holds ``for every $\eta\in[-1,1]$'', and its
item (v) requires $S(\infty,\eta)=0$; a profile with $U$ odd in $\eta$ has $U(X,0)=0$, so
$S(\infty,0)=-\int_0^\infty E(X,0)^2/2\,dX<0$ whatever the amplitude, $-1.1$ at $c_\infty=0.2$ and $-4.1$ at $0.4$,
which are the residuals observed. The leading profile of the \OAI{} construction therefore carries axial velocity on
the dividing plane: the core is an axial through-flow, spun up by the radial inflow and evacuated along the axis
in one direction, with the return flow at larger radius so that $M(\infty)=\int_0^\infty U\,dX=0$. It is not the
$z$-symmetric core of the \GD{} problem, which all of our earlier steps had assumed, and it is what \OAI{}
says of itself in its Section~2.1: with exact reflection symmetry the axial transport of angular momentum
would vanish at $z=0$ and so would $u_z(r,0,t)$, leaving no radial shear of the axial velocity to amplify the
pulses near the middle plane, so the authors ``choose a slightly asymmetric axial profile, with a small upward
bias and nonzero velocity at $z=0$''. Our matching functional recovers that choice from the identities alone.
Dropping the symmetry opens the matching, and freeing the annulus content closes it at $c_\infty=0.2$. Over
$F_0,U_0\in{\rm span}(T_0,T_1,T_2)$ and an annulus return flow $(b_0+b_1\eta)\psi(X)$, eight parameters, the
residual falls from $0.36$ to $0.077$ ($0.039$ with thirteen parameters) against natural scales of order one, with
the matched data $F_0=0.244+0.034\eta-0.058T_2$, $U_0=0.450+0.004\eta-0.034T_2$ and $b_0=-2.01$; the optimum
moves by less than $5\cdot10^{-3}$ between $N_\eta=32$ with 24 modes and $N_\eta=24$ with 20, and the core identity
holds to $10^{-6}$ on it. The annulus content is the freedom \OAI{} itself uses: it does not continue the core
through the annulus but chooses the profile there piecewise (its Prop.~A.4 and Lemma~A.5) and lets the pulse stress
make up the difference. Writing that content as radial and axial Chebyshev modes under a $C^\infty$ bump, with the
axial part compact in $(X_a,X_v)$, $X_v\le X_b$, as Theorem~4.6(v) asks of the axial pulse, and taking $F_0,U_0$ to
$T_4$, 32 parameters in all, the residual falls to $1.2\cdot10^{-3}$ root-mean-square and $3.7\cdot10^{-3}$ at worst,
$0.1\%$ of the natural scale, with no identity dominant and a flat residual spectrum in $\eta$. Re-evaluated at
$N_\eta=40$ with 28 modes the same data give $2.0\cdot10^{-3}$ and $1.2\cdot10^{-2}$, the excess in the net-force
identity at the singular ends, while the profile, its pressure deficit and the core identity are unchanged to five
digits: the match holds to $0.2\%$ across resolutions, the last factor of two being the $\eta$ truncation, and what
remains is the truncation of the family and of the filter, not an obstruction; re-optimized at $N_\eta=40$ the residual settles at $1.6\cdot10^{-3}$. The matched core has $F_0=0.336$ and $U_0=0.452$ on the dividing
plane and $F_0=0.214$ at $\eta=1$, an annulus return flow led by $-2.06\,\psi(X)$ and an annulus swirl deficit
growing outward, $|U|$ reaching $0.60$ in the core and $1.9$ in the annulus, the core identity at $10^{-6}$ and the
pressure datum consistent to $5\cdot10^{-13}$. At $c_\infty=0.3$ and $0.4$ the same family gains nothing ($0.17$ and
$0.37$), and the reason is now located. The pressure datum solves a fixed-point equation whose fixed point is
repelling; Anderson mixing reaches it, and so does a Newton iteration on its Chebyshev modes, to the same datum
within $10^{-9}$, and both stall at the same place, which is therefore a property of the equation: along a family
of axis data with growing through-flow, $U_0\to sU_0$, the residual of the fixed-point equation jumps from
$10^{-13}$ to $10^{-6}$ and grows beyond a critical $s$, a fold at which $I-G'$ is singular along the Newton
direction and beyond which no consistent datum exists for the smooth-blend core. At $c_\infty=0.2$ it sits at
$U_0(0)=0.835$ and at $0.4$ at $0.925$ (at $0.3$ the jump is only to $10^{-8}$ and the fold is marginal), and its
location depends on the whole datum, since the $c_\infty=0.2$ axis data keep a consistent datum at $c_\infty=0.6$
over the whole range tried. The least squares at $c_\infty\ge0.3$ sits against this fold with $U_0(0)=0.75$ and
$0.93$, and the amplitude the $S$ identity asks for is carried by the annulus return flow ($-2.9$ and $-3.4\,\psi$,
$|U|$ up to $3.9$ there) rather than by the core. The cone fails on every matched profile ($\kappa\le-10^3$,
admissible on at most $7\%$ of the annulus): \OAI{} enforces the cone by the radial oscillation with phase
$N\log X$ of its Prop.~C.2, which changes the shear at $O(1)$ while moving the profile and its moments by $O(1/N)$,
and a smooth profile has no such freedom. The cone requirement has a physical reading that the next step rests on. With circulation $\Gamma=ru_\theta$, the profile variables give $d\log\Gamma/d\log X=1-a/2$, so $a>2$ is Rayleigh's criterion for centrifugal instability, circulation decreasing outward, and $v_s=a(1+t_s^2)>2$ in Theorem~4.6(iii) of \OAI{} is its form with the axial shear included, Ludwieg's criterion for a swirling flow with axial shear: the annulus must be centrifugally unstable, and the pulses are that instability fed by the shear. Our matched profiles have $a<0$ over part of the annulus, where the swirl still grows outward: centrifugally stable, with no free energy for the pulses, which is why the cone fails. The swirl alone cannot do it everywhere. On the dividing plane the leading term of $p_{s,1}$ is $G=I/H-X$ with $I=\int_0^XH\,dx$, the exterior fixes $G$ at $X_b$, and $dG/dX=(I/H)(a/2-1)/X$, so positive stress inside the annulus requires $G$ to fall toward $X_b$, that is $a<2$ near the outer edge; there $v_s>2$ must come from $t_s$, the radial shear of the axial velocity, which is what Section~2.1 of \OAI{} says supplies the amplification near the middle plane. The heat exterior itself sits exactly on the boundary of the cone, $P_c=v_s=2+2h$ with vanishing stress, so any margin must vanish at the outer edge. We then built the annulus the way \OAI{} does (its Section~A.2 and Proposition~A.4), at finite size: a smooth join of the marched core onto a prescribed swirl decaying as a power law $H\propto X^{-\lambda}$, an axial pulse of three lobes whose amplitudes set $M(\infty)=J(\infty)=S(\infty)=0$ pointwise in $\eta$, an interpolation of the swirl onto the heat exterior, two swirl bumps for the angular-momentum identity with the pressure integral preserved, and $c_\infty$ from the mid-plane identity. With the join at $X_a=2$, $\lambda=0.5$ and odd axis data $U_0=0.3\eta$ this meets all four identities to $10^{-13}$ with pulse amplitude $5.6$ and $|U|\le2.1$ in the annulus: the matching problem is solved in the construction's own manner, without any least squares. The cone then holds in the quiet stages, where $a=v_s=2+2\lambda=3$, and fails in the pulse, for a reason that scales: the pulse's radial shear is $b_s=2XU_X/E\sim2\,{\rm Amp}/\Delta y$ for a pulse of logarithmic width $\Delta y$, so $v_s=a(1+t_s^2)$ reaches $10^2$ to $10^4$ while $P_c\sim XQ_s$ is of order $10$. Admissibility $P_c>v_s$ needs $XQ_s\gtrsim4\,{\rm Amp}^2/(a\,\Delta y^2)$: with the amplitude fixed by the $S$ deficit, either $X$ of order $10^2$--$10^3$ for a pulse of unit width or $\Delta y$ of order $10$ at $X$ of order $10$. Neither can be reached with a core computed as a Cauchy problem from the axis, which cannot be continued past $X\approx4$; both are exactly what Proposition~A.4 prescribes, a reference inner power law out to a large radius $X_R$ and a pulse of length $13/\lambda$. We therefore built that version too: the computed core kept to $X=2.5$ and blended onto the construction's reference inner profile (its (A.7): $E\propto(1+\eta^2)^{-1}X^{1/10}$, $U\propto\eta$) out to $X_R=50$, then the stages of its Section~A.2 with the terminal dip of (A.12) closing the angular-momentum identity. It reproduces the construction stage by stage. The four identities hold to $10^{-10}$; the swirl slopes are exactly the prescribed $0.8$, $2$, $2+2\lambda$ and $2+2h$; the stress is positive from $X\approx14$ outward; on the reference interval the axial through-flow drives the radial-velocity term $W$ negative and lifts $P_c=XQ_s/L$ to $20$--$35$ with a margin of $6$--$20$ in the relaxed cone, which is the mechanism Proposition~A.4 relies on and the reason the construction wants an axial flow there; and on the power-law stage the admissible cone holds outright. The pulse and terminal stages do not close, for two reasons that are now quantitative. Inside the pulse $P_c=X(Q_s+t_sN_s/E)/L$ changes sign with the pulse's slope, since $t_sN_s/E\approx-\tfrac23R_bR_b'$ for $U=ER_b$, so admissibility needs $R_bR_b'<Q_s\approx\lambda/(1-\lambda)$; and the pulse must supply the swirl's own kinetic-energy deficit $\int E^2/2$ over the whole annulus, so $R_b^2\approx Y/(2\Delta y)$ for a pulse of log-width $\Delta y$ in an annulus of log-length $Y$. Together these give $\Delta y\gtrsim13/\lambda$: the construction's pulse length, derived here from its energy budget. At the radii that length requires, $X\sim10^{20}$, the axial stress $T_{0,z}=F(p_{s,2}+b_s)$ with $p_{s,2}=XN_s/(LE)$ amplifies any residual of the pressure and $S$ integrals by $X$; the construction cancels those residuals identically, stage by stage, and a computation at $N_\eta=24$ with quadrature leaves one part in $10^6$, which is astronomical after that amplification. The cone condition is therefore a statement about the proof's asymptotic regime: consistent, computable stage by stage, and not a property that any profile of order-unity logarithmic extent in the similarity radius can have. What these steps establish is the machinery, verified against the
residual route, the through-flow character of the core, the amplitude scale $c_\infty/\sqrt h$ that the $S$
identity sets, a matched profile at the smallest amplitude, and the fold of the pressure datum that bounds the
smooth-blend core. Closing at larger amplitude needs a core representation whose datum exists there, the
piecewise core of \OAI{} rather than a smooth blend. With the 32-parameter annulus the closure degrades smoothly with amplitude, root-mean-square $0.0016$, $0.0022$, $0.0037$ and $0.059$ at $c_\infty=0.20$, $0.22$, $0.25$ and $0.30$, and $c_\infty=0.4$ stays at $0.37$; the fold of the pressure datum sits at a swirl amplitude only $23\%$ above the closed match's, at an axis through-flow $1.85$ times it, and is not reached by scaling the odd part of the axial data or the annulus content alone.

\begin{figure}[!t]\centering
\includegraphics[width=0.9\textwidth]{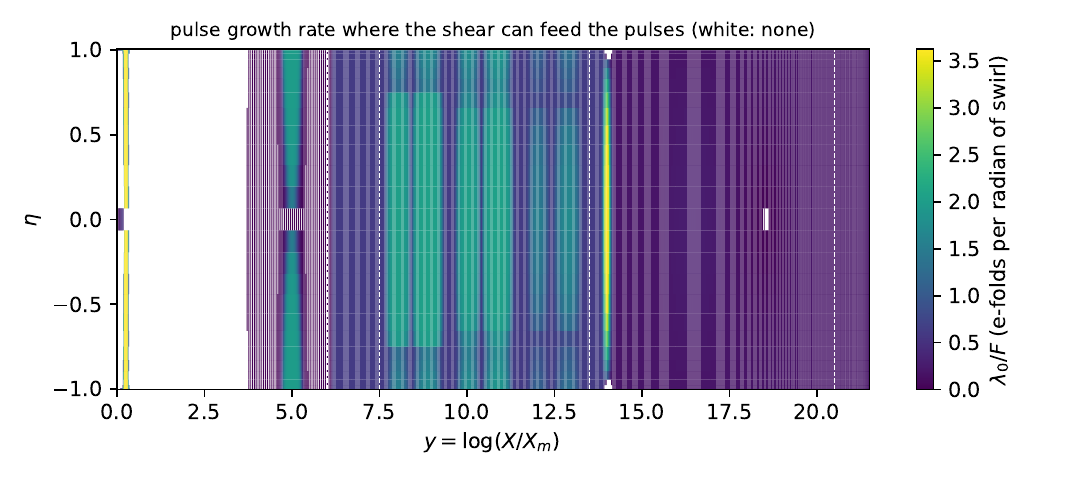}
\caption{Growth rate of the pulses of \OAI, $\lambda_0/F=\sqrt{2a(1-2/v_s)}$ in e-folds per radian of swirl, on the large-radius construction of Section~\ref{sec:axis} ($\lambda=0.1$), against $y=\log(X/X_m)$ and $\eta$; dashed lines mark the stage boundaries (reference interval, axial shut-off, power law, pulse, interpolation and terminal dip). White: the shear cannot feed the pulses ($v_s\le2$); on the reference interval this is the case, as in Proposition~A.4 of \OAI, and the modulation of its Proposition~C.2 is what supplies the missing shear there.}\label{fig:pulse}
\end{figure}

\paragraph{Pulse growth.} The pulses of \OAI{} are transverse waves fed by the radial shear of the tangential velocity, and their amplitude equation (its (7.5) and Lemma~7.1) gives their growth rate in closed form in the profile variables: with $g_0=F(-a,b_s)$ the shear vector, the rate is $\lambda_0=F\sqrt{2a(1-2/v_s)}$, real exactly where $a>0$ and $v_s>2$, so the Ludwieg condition of the cone is the condition that the pulses can grow at all, and the ratio of tangential to radial amplitude is $c_0=-\sqrt{(v_s-2)/2}$. Since $F$ is the angular velocity in similarity units, $\lambda_0/F$ is the number of e-folds per radian of the core's rotation. Along a pulse the wavevector tilts with the shear, $s(v)=u_*(\tfrac12+v/L_s)$, the rate falls to $\lambda_0/\sqrt{1+s^2}$ while viscous damping rises as $\lambda_0(1+s^2)/(1+u_*^2)^{3/2}$, and the envelope peaks where the two cross; the tilt $u_*$ is fixed by the cone margin through $u_*/\sqrt{1+u_*^2}>|J_c|/(P_c-v_s)$. On a power law $H\propto X^{-\lambda}$ with no axial shear the formula reduces to $\lambda_0/F=2\sqrt\lambda$, and on the heat exterior to $2\sqrt h=0.2$: the anomalous exponent sets the pulses' growth in the far field. We evaluated $\lambda_0/F$ on the computed profiles (Figure~\ref{fig:pulse}). On the smooth matched profile at $c_\infty=0.2$ the shear can feed pulses on $47\%$ of the annulus, with a median rate of $2.4$ e-folds per radian where it can, and none where $v_s\le2$. On the large-radius construction the reference interval feeds none ($v_s=a=0.8$), which is why its Proposition~C.2 loop is needed there; the power-law stage gives $0.63=2\sqrt{0.1}$; and inside the pulse the axial shear raises the rate to $0.6$--$2.1$ e-folds per radian. For the experiment this is the instability to seed: an azimuthal wave at the edge of the core, growing by a factor $e$ every $0.5$ to $5$ radians of swirl, fastest where the axial through-flow shears the swirl.

\paragraph{Axis pressure deficit.} One number carries into Section~\ref{sec:fluid}. Integrated to infinity by the
eq.~(4.25) of \OAI, $\Pi(X,\eta)=-\int_X^\infty E^2/(2x)\,dx$, the deficit $\Pi(0,0)-\Pi(\infty,0)$ is $0.77\,u_{\max}^2$
for the sample data (the value $0.48$ first obtained was truncated at $X=3$), $0.56$ for the pinned fit and $0.58$
in the small-amplitude limit, $u_{\max}$ being the peak swirl proxy $\sqrt{2X}F$ on the dividing plane; for the
matched through-flow profiles the coefficient is $0.34$--$0.61$ ($0.34$ for the closed match at $c_\infty=0.2$,
$0.50$ for the eight- and eighteen-parameter fits at the same amplitude, $0.54$ and $0.61$ at $c_\infty=0.3$ and $0.4$;
symmetric and unmatched profiles gave $0.56$--$1.3$), so the cavitation threshold in water, deficit equal to one
atmosphere, is reached at $u_{\max}$ between $12.8$ and $17.1$~m/s. The low end carries a caveat: in the closed
match the annulus swirl content moves the swirl peak into the annulus, at $X=1.3$, where the moment identities
constrain the profile only through its integrals, so the coefficient is a property of the matched family, and the
piecewise core of \OAI{} may give another. The axial velocity on the axis is $0.5$ to $1.2$ times the peak swirl on
every matched profile: the axial jet is of the same order as the swirl. For axis data of order one
the consistent pressure datum drives the axial velocity so hard that the series radius collapses below the join;
those cases need the march.

\section{What a fluid would do with this singularity}\label{sec:fluid}
This section serves question (b) raised in our abstract. Everything in it follows from the similarity scalings of Section~\ref{sec:2026} and is an order-of-magnitude estimate; the computed profiles that would turn the estimates into inception times and radii require the axis core of Section~\ref{sec:axis} and are marked as such.

\paragraph{Scalings.} The blow-up is energetically free, and its anomalous factor $\tau^{-h}$ is $1.4$ at $\tau=10^{-15}$. On the dividing plane the collapse scale is the time to the singularity, $q\sim\tau$; the core radius scales as $r\sim\sqrt q$, its
height as $z\sim q^D$, and the velocities as $q^{-A}$ by \eqref{eq:smooth}. Table~\ref{tab:scalings} lists
the consequences. Two of them matter. The kinetic energy in the core vanishes as $\tau^{1/2-3h}$ while the
dissipation rate diverges as $\tau^{-1/2-3h}$ with a finite time integral: the blow-up is energetically free,
which is the physical content of the bounded $L^2$ norm in the theorem. And the anomalous factor $\tau^{-h}$
that distinguishes the \OAI{} construction from an ordinary viscous swirl collapse ($h=0$, for which $u_\theta r$
is conserved and $\sup|\omega|$ has an integrable time singularity) is $10^{0.15}=1.4$ at $\tau=10^{-15}$ for
$h=1/100$; any physical cutoff arrives while the flow is, to a few percent, indistinguishable from the
ordinary collapse.

\begin{table}[h]\centering\small
\begin{tabular}{lll}\toprule
quantity & scaling in $\tau$ & behavior as $\tau\to0$ \\\midrule
peak swirl velocity & $\tau^{-1/2-h}$ & diverges \\
core radius; core height & $\tau^{1/2}$; $\tau^{1/2-h}$ & vanish \\
kinetic energy in the core & $\tau^{1/2-3h}$ & vanishes \\
peak vorticity & $\tau^{-1-h}$ & diverges \\
dissipation rate in the core & $\tau^{-1/2-3h}$ & diverges; time integral finite \\
$\int_\tau^1\sup|\omega|\,d\tau'$ (Beale--Kato--Majda) & $(\tau^{-h}-1)/h$ & diverges, only just \\
axis pressure deficit $\sim\rho u_\theta^2$ & $\tau^{-1-2h}$ & diverges \\\bottomrule
\end{tabular}
\caption{Similarity scalings of the core on the dividing plane, from $r\sim\sqrt q$, $z\sim q^D$,
$u\sim q^{-A}$ and $q\sim\tau$, with $A=\tfrac12+h$, $D=\tfrac12-h$. For $h=1/100$ the anomalous factor
$\tau^{-h}$ is $1.05$ at $\tau=10^{-2}$, $1.12$ at $10^{-5}$ and $1.41$ at $10^{-15}$.}\label{tab:scalings}
\end{table}

\paragraph{Which continuum assumption fails first.} In a liquid it is cavitation, at a core radius near a millimeter for laboratory scales; in a gas it is compressibility, near thirty micrometers; molecular scales are never reached. Take water, an initial swirl of $1$~m/s at a radius of
$1$~cm, and follow the collapse with $u_\theta r\approx$ const (the $\tau^{-h}$ factor is within $5\%$ of
unity over the range that matters). The axis pressure deficit $\rho u_\theta^2$ reaches one atmosphere,
$10^5$~Pa, when $u_\theta^2=10^5/10^3$~m$^2$/s$^2$, i.e.\ $u_\theta\approx10$~m/s ($12.8$--$17.1$~m/s with the
computed deficit coefficients of Section~\ref{sec:axis}), and $15$~m/s if the local
pressure must fall a further atmosphere below vapor pressure before nuclei grow. That is a core radius of
$0.6$--$1$~mm ($0.59$--$0.78$~mm on the matched profiles) and $\tau/\tau_0=(r/r_0)^2\approx3\cdot10^{-3}$ to $10^{-2}$: vortex-core cavitation inception, the mechanism of
tip-vortex cavitation reviewed by Arndt \cite{arndt02} and studied at Dynaflow \cite{chahine04}. Viscous heating is negligible there: the
dissipation per unit mass is $\nu(u_\theta/r)^2\approx10^{-6}\times(10^4)^2=10^2$~W/kg, the remaining
collapse time on the viscous scale $r^2/(2\nu X)$ is of order $0.5$~s, so the temperature rise is of order
$50$~J/kg$\,/\,4\cdot10^3$~J/(kg\,K)~$\approx0.01$~K. Molecular scales ($0.3$~nm) lie eight decades below
the initial radius and seven below the inception radius, thirteen decades further in $\tau$. In air the
first failure is incompressibility: $u_\theta$ reaches the speed of sound, $340$~m/s, at a velocity factor of
$340$, a core radius of about $30$~$\mu$m and $\tau/\tau_0\approx10^{-5}$; the Knudsen number
$\lambda/r$ with $\lambda=68$~nm reaches $10^{-2}$ at $r\approx7$~$\mu$m and $10^{-1}$ at $0.7$~$\mu$m, one
and three decades later in $\tau$. Physically realized, the forced blow-up is a cavitating vortex in a
liquid and a shocking one in a gas, arrested while $\tau^{-h}$ is $1.05$ (water) and $1.12$ (air). None of
this can be supplied by the annulus problem, which has no axis. The matched cores of Section~\ref{sec:axis} turn
the estimate into a number with a stated spread, $u_{\max}=12.8$--$17.1$~m/s and a core radius of $0.6$--$0.8$~mm for
the initial swirl above, with the caveat that the coefficient depends on the annulus content; the cone condition of
the \OAI{} construction decides whether the forced annulus can be built around the core at all, and on the smooth
profiles of Section~\ref{sec:axis} it fails everywhere, so the annulus needs the radial oscillation of \OAI, and the
inception numbers are those of the leading-order profile, not of the forced flow.

\paragraph{The experiment one could propose.} The apparatus is the \GD{} chamber and the observables are axis cavitation inception and the loss of the steady state at the fold. A swirl chamber with porous walls and controlled filtration,
the \GD{} Dynaflow geometry \cite{gd98,chahine96}, driven toward collapse by increasing the inflow at fixed
swirl or the swirl at fixed inflow, with cavitation inception on the axis as the primary observable, an
acoustic event with a well-developed detection art. The sweep supplies a second observable that does not
require the axis: at an inflow Reynolds number of $10$ the steady symmetric swirl between the porous walls
ceases to exist beyond a wall swirl of $\Fhi=40.47$, azimuthal Reynolds number $81$ at the reference instant
(Section~\ref{sec:layer}), a fold converged on three grids, with a hysteresis window $[37.3,40.5]$ below it.
Loss of the steady state at a critical swirl at fixed inflow, and its return at a lower swirl, is a
measurement a torque meter or a swirl probe can make. The critical swirl at other inflows follows the fold law
\eqref{eq:foldlaw} and Table~\ref{tab:fold}: in the experiment's terms the steady swirl is lost at an azimuthal
Reynolds number of $0.8$ to $0.85$ times the square of the radial one, from $66$ at radial Reynolds number $9$
to $341$ at $20$. The state that replaces the steady one beyond the fold is not computed (Section~\ref{sec:layer}),
and the inception radii on the axis are those of the matched core at the smallest exterior amplitude
($0.6$--$0.8$~mm from $1$~cm at $1$~m/s, Section~\ref{sec:axis}); the first is open, the second awaits the match at
larger amplitude.

\section{Discussion}\label{sec:disc}
\paragraph{What is established.} On question (a), computability, the answer is yes for the leading-order operators of the \OAI{} construction: they can be discretized and solved to spectral accuracy, and a self-similar collapsing swirl exists between porous walls held at fixed similarity radii over the whole range of the \GD{} parameters and well beyond it. Below inflow $9$ it is unique in that class as far as it was traced, reflection-symmetric, and responds to a symmetry-breaking datum smoothly, with a linear response that peaks near unit-order swirl and reverses sign at larger swirl; above inflow $9$ the symmetric branch folds, at a swirl $F^*_{\rm hi}\simeq0.41V_0^2$ converged on three grids, and the fold is a saddle-node of the collapse dynamics, with the symmetry-breaking mode running ahead of it. The profile is an attractor of the collapse at weak inflow and along the whole approach to the fold. On the axis, the Dirichlet problem has no resolution-stable solution, so the core must be computed as a Cauchy problem in $X$ from analytic axis data, for which \eqref{eq:R1}--\eqref{eq:R2} is the recursion, and that problem returns one fact about the flow: the moment identities of \OAI{} cannot be met by a core symmetric about the dividing plane, so the blow-up core is an axial through-flow with a return flow at larger radius, as its authors chose it; a non-symmetric core with free annulus content meets the identities to $0.2\%$ at the smallest exterior amplitude tried. The cone condition of \OAI{} is Rayleigh's centrifugal criterion with axial shear, and built the construction's own piecewise way it requires similarity radii of order $10^{20}$: it is a statement about the asymptotic regime of the proof, not about a flow one could compute or build. On question (b), the scalings say that the blow-up is energetically free and that any real fluid leaves the continuum description, by cavitation or by compressibility, while the anomalous factor $\tau^{-h}$ is within $12\%$ of unity; a laboratory realization is a viscous swirl collapse arrested by inception or by a shock, and the \GD{} chamber driven toward collapse is its apparatus, with inception on the axis, the loss of the steady state at the fold and a one-directional axial jet as observables. Put together: the leading-order flow of the \OAI{} construction can be computed, and its core is the axial through-flow its authors chose; the forced annulus that sustains it cannot be built at any radius a computation reaches, the cone condition placing it at similarity radii of order $10^{20}$; in the porous-wall model that served as the testbed, the state beyond the fold and the spectrum at moderate inflow with strong swirl are not converged, and resolving them is a formulation problem rather than a clue about blow-up; and long before the singular time a real fluid leaves the incompressible description, water by cavitation and air by compressibility, while the anomalous factor is still within $12\%$ of one. Nothing found here suggests that the mechanism is reachable in a flow one computes or builds, and the forced theorem says nothing about the unforced equations. Those who solve the Navier--Stokes equations in engineering practice can continue to rely on them; on the unforced question this study is silent, as the theorem is.

\paragraph{What was learned about method.} The \GD{} solver's skeleton, Chebyshev collocation and continuation, carried over with three changes: Newton in place of successive substitution, an explicit pressure gauge, and pseudo-arclength in place of relaxation. Four rules were found on the way, each by being caught out by it. The truncated $\eta$ range is admissible only under \eqref{eq:rule}; violating it turns an outflow boundary into an inflow one and manufactures bifurcations that pass every local test. The inner-wall layer at strong inflow wants a clustered radial grid; 32 mapped modes then do what 64 plain ones cannot. The Cauchy problem in $X$ is ill posed with a measured growth law and is made usable by a mode filter, not by resolution. And where the axial transport has an interior sonic line with $|U|\gtrsim30$ on it, at moderate inflow and strong swirl, the collocation in $\eta$ converges erratically, the leading eigenvalues wander between grids that all resolve the layer, and no amount of resolution settles the spectrum; the same line is present at strong inflow, where everything converges, so the presence of the line is necessary but not sufficient and the mechanism is not understood. The heat exterior of \OAI{}, absent from \GD{}, entered at every stage: as the verification solution, as the far field of the join and the normalization of the matching, as the source of the algebraic $\eta$ convergence, and as the tail term that forces the through-flow.

\paragraph{What is open.} Six items, each with the step that would settle it.
\begin{enumerate}
\item The profile problem at moderate inflow and strong swirl (Section~\ref{sec:stab}). A formulation that imposes the analytic branch across the interior sonic line, or restores the $O(q^{2h})$ axial viscosity in a layer about it, and a re-run of the sweep and the spectrum there. This is a question in the numerical analysis of degenerate first-order problems before it is one in fluid mechanics.
\item The sheet beyond the fold (Section~\ref{sec:layer}), which no symmetric collocation converged. The same formulation, or a time-dependent computation in the rescaled variables, would show what replaces the steady state.
\item The core at larger exterior amplitude (Section~\ref{sec:axis}). The closure degrades from $0.2\%$ at $c_\infty=0.2$ to $6\%$ at $0.3$ and the pressure datum of the smooth-blend core ceases to exist at a fold; the piecewise core of \OAI{} in place of the smooth blend is the next step, and it is the authors of \OAI{} who know that object best.
\item The cone at finite radius. The large-radius construction reproduces the mechanisms of \OAI{} stage by stage but needs $X\sim10^{20}$ and cancellations beyond double precision; the modulation of Proposition C.2 of \OAI{}, or an asymptotic matching in place of a computation, is what would close it.
\item The stability of the axis core. The dynamic-rescaling machinery of Section~\ref{sec:stab} exists for the annulus; applying it to the Cauchy-in-$X$ core requires a treatment of the ill-posed radial direction that was not attempted.
\item The experiment (Section~\ref{sec:fluid}), and before it a forced Navier--Stokes computation at finite viscosity. Both are for a fluids laboratory and a computing group that work in this area.
\end{enumerate}

\paragraph{Handoff.} This is where \RD{} leaves the problem. The two questions that started it are answered above; what remains is listed with a first step for each, and the material to take those steps is released with the paper, at \url{https://gitlab.umiacs.umd.edu/ramanid/swirl-collapse} and as ancillary files: the solver, some six thousand lines of Python (tensor Chebyshev collocation, complex-step Newton, arclength continuation, the axis series and march, the join, the matching functional, the dynamic-rescaling march and eigenproblem), its test suites, the reports on each stage, and the research log that records every result and every failure with its reason. Appendix~\ref{app:methods} is a short tutorial map from the methods to the code, written for whoever takes an item up. The run outputs themselves are available from \RD{} on request. Anyone who takes up an item, or finds an error, is asked to write.

\section*{Acknowledgments}
In memory of Nail A. Gumerov (d.\ 2022), co-author of \GD{}, the 1998 work that this paper builds on.
Computations were performed on a laptop and on the Zaratan cluster of the University of Maryland (Division of
Information Technology) and the Nexus cluster of UMIACS; every result was verified by the tests of Section~\ref{sec:verif}.
We thank the staff of both facilities. The role of Claude Fable 5.1 in this work is described in the statement on generative AI on page~2.

\appendix
\section{The numerical methods, briefly, with pointers to the code}\label{app:methods}
This appendix is for whoever takes up an item of Section~\ref{sec:disc}. Each paragraph says what a method is in a few sentences, why it was used here, what it caught us out on, where it lives in the code at \url{https://gitlab.umiacs.umd.edu/ramanid/swirl-collapse}, and where to read more. File and function names are in typewriter; every script has a docstring at the top that says what it does and which report in \texttt{results/} discusses its output. Table~\ref{tab:map} maps the sections of the paper to the scripts and the reports.

\begin{table}[!htb]\centering\small
\setlength{\arrayrulewidth}{0.3pt}
\begin{tabular}{p{0.30\textwidth}|p{0.36\textwidth}|p{0.28\textwidth}}
\hline
section of the paper & scripts & report \\ \hline
\ref{sec:2026}, \ref{sec:1998} operators and dictionary & \texttt{core\_solver.py} (docstring and \texttt{Grid.T}, \texttt{Grid.Z}), \texttt{tests.py} & \texttt{notes.tex}, \texttt{results/PAPER\_UPDATES.md} \\
\ref{sec:num} method & \texttt{core\_solver.py}, \texttt{continuation.py}, \texttt{deflation.py} & -- \\
\ref{sec:verif} verification & \texttt{tests.py}, \texttt{tests\_phase0.py}, \texttt{line0.py} & \texttt{results/line0\_summary.md}, \texttt{results/replicas.md} \\
\ref{sec:sweep} sweep, rules, fold & \texttt{sweep.py}, \texttt{run\_annulus.py}, \texttt{fold\_locus.py}, \texttt{pitchfork\_locus.py}, \texttt{asym.py}, \texttt{branch*.py}, \texttt{*\_program.py}, \texttt{deflation.py} & \texttt{results/sweep\_summary.md}, \texttt{folds.md}, \texttt{deflation*.md}, \texttt{line5\_slopes*.md}, \texttt{FINDINGS\_live.md} \\
\ref{sec:stab} stability & \texttt{annulus\_stability.py}, \texttt{tests\_stability.py} & \texttt{results/stability\_report.md} \\
\ref{sec:axis} axis core & \texttt{axis\_series.py}, \texttt{axis\_march.py}, \texttt{axis\_join.py}, \texttt{axis\_march\_match.py}, \texttt{axis\_pi0\_fold.py}, \texttt{axis\_cone.py}, \texttt{axis\_cone\_loop.py}, \texttt{axis\_piecewise.py}, \texttt{axis\_pulse\_growth.py}; tests \texttt{tests\_axis.py}, \texttt{tests\_march.py}, \texttt{tests\_join.py}, \texttt{tests\_cone.py} & \texttt{results/axis\_series\_report.md}, \texttt{axis\_march\_report.md}, \texttt{axis\_join\_report.md}, \texttt{axis\_cone\_report.md}, \texttt{axis\_step5\_report.md}, \texttt{axis\_piecewise\_report.md} \\
\ref{sec:fluid} fluid and experiment & \texttt{axis\_inception.py} & \texttt{results/axis\_step5\_report.md} \\
figures & \texttt{paper/make\_figures.py}, \texttt{make\_hero.py}, \texttt{make\_stability\_fig.py}, \texttt{make\_boundary\_fig.py} & -- \\
clusters & \texttt{cluster/*.sbatch}, run lists \texttt{cluster/*.txt}, \texttt{cluster/status\_dash.py} & \texttt{HANDOFF.md} \\ \hline
\end{tabular}
\caption{Where each part of the paper is computed and discussed.}\label{tab:map}
\end{table}

\paragraph{Chebyshev collocation in two variables.} A smooth function on $[-1,1]$ is represented by its values at the Gauss--Lobatto points $x_j=\cos(\pi j/N)$, and derivatives are taken by multiplying the vector of values by the differentiation matrix $D$; for analytic functions the error decays geometrically in $N$, for functions with a singularity on the interval only algebraically. Here the unknowns $F$, $U$ live on a tensor grid in $(X,\eta)$, $X$-derivatives act on the first index and $\eta$-derivatives on the second, and the operators $T_b$ and $Z_b$ of Lemma~4.1 of \OAI{} are formed exactly this way (\texttt{core\_solver.py}: \texttt{cheb}, \texttt{cheb\_interval}, class \texttt{Grid} with \texttt{dX}, \texttt{dE}, \texttt{T}, \texttt{Z}). Two things to know: the number of $\eta$ nodes must be even so that a node sits at $\eta=0$, where the gauge is pinned; and the first-order $\eta$ direction takes no boundary condition at the outflow ends, so the $\theta$ and $z$ equations are collocated at every $\eta$ node, including $\pm1$. Trefethen \cite{trefethen00} is the shortest introduction, Boyd \cite{boyd} the most complete on pitfalls, Canuto et al.\ \cite{canuto} the reference.

\paragraph{Integration matrices and the pressure gauge.} The pressure is eliminated through $\Pi=\Pi_0(\eta)+\int_{X_{\rm lo}}^X F^2\,dx$ and the radial velocity through continuity, both by a spectral integration matrix $Q$ with $Q D=I$ on functions vanishing at the lower end (\texttt{cheb\_integration\_matrix}, \texttt{Grid.intX}). The datum $\Pi_0(\eta)$ is an unknown determined by the flux condition at the outer wall, and the discrete Jacobian then has one exact null direction, a constant shift of $\Pi_0$; it is removed by replacing the flux equation at the mid-plane node by $\Pi_0(0)=0$ (\texttt{Problem.residual}, \texttt{gauge\_fix}, \texttt{j\_pin}). Forgetting the gauge does not stop Newton from converging on a symmetric grid, it makes the Jacobian singular by one and every continuation step ill-conditioned; the compatibility residual of the replaced equation is kept and checked (\texttt{compat\_residual}).

\paragraph{Mapped grids.} When the solution has a thin layer, more nodes near it are worth more than more nodes overall. The radial grid can be clustered at either wall by a $\sinh$ map, $X(s)$ with $X'(s)>0$ everywhere so that $d/dX=(1/X')\,d/ds$ (\texttt{mapped\_x\_grid}; parameter \texttt{xmap}, negative for the inner wall); the axial grid can be clustered at both ends by $\eta=\tanh(as)/\tanh a$ (\texttt{Grid} with \texttt{emap}). The inner-wall layer at strong inflow is resolved by 32 mapped modes and not by 64 plain ones (Section~\ref{sec:layer}); the sonic layer of Section~\ref{sec:stab} is resolved by the $\tanh$ map in the sense of node count, and the eigenvalues still do not converge, which is the open item. Bayliss and Turkel \cite{bayliss92} on mappings for layers; the maps here are the simplest that keep $X'(s)$ bounded away from zero.

\paragraph{Complex-step Jacobian and Newton.} The residual is evaluated in complex arithmetic at $y+i\varepsilon e_k$ with $\varepsilon=10^{-30}$ and the imaginary part divided by $\varepsilon$ is the $k$-th Jacobian column to machine precision, with no cancellation error, provided the code contains no operation that is not analytic in its arguments (\texttt{Problem.jacobian}; \texttt{continuation.complex\_step\_jacobian}). It costs one residual evaluation per unknown, which is why the runs of Section~\ref{sec:stab} at $3000$ unknowns take minutes per Newton step; it was chosen over hand-coded derivatives because the residual changed a dozen times in four days and never had to be re-differentiated. Newton is damped by backtracking (\texttt{Problem.newton}) and, in the continuation of Section~\ref{sec:stab}, restricted to the reflection-symmetric subspace when it fails near a symmetry-breaking point (\texttt{annulus\_stability.newton\_sym}). The round-off floor of the residual grows like $N_\eta^2$ and exceeds $10^{-10}$ beyond about $30$ nodes, so convergence there is declared at stagnation below $10^{-7}$ rather than at a fixed tolerance. Squire and Trapp \cite{squire98}, Martins et al.\ \cite{martins03}.

\paragraph{Pseudo-arclength continuation and fold detection.} A solution branch $y(\lambda)$ that folds cannot be followed in $\lambda$; one parametrizes it by arclength instead, adds the equation that the step be of prescribed length along the tangent, and solves the bordered system, which is regular at folds (\texttt{continuation.ArclengthContinuation}, options in \texttt{ArclengthOptions}). The sign of the determinant of the bordered Jacobian, from the LU factorization, changes at folds and at pitchforks (\texttt{lu\_det\_sign}), and this is how the fold locus and the determinant sign changes of Section~\ref{sec:sweep} were recorded. The lesson of Section~\ref{sec:stab} is that a determinant sign change on one grid is a claim about that grid: the ones at inflow $7$ and $8$ did not survive refinement. Keller \cite{keller77}, Allgower and Georg \cite{allgower90}.

\paragraph{Deflated Newton.} To find out whether a problem has more than one solution, divide the residual by a factor that blows up at each solution already found; Newton then cannot return to them and either finds a new one or fails (\texttt{deflation.deflated\_newton}, \texttt{initial\_guesses}). Used in Section~\ref{sec:axis} to show that the Dirichlet axis problem has no resolution-stable solution: deflation finds a different solution on every grid and none of them persists. Farrell, Birkisson and Funke \cite{fbf15}.

\paragraph{Verification.} Three independent checks, all in \texttt{tests.py}: the operators are derived symbolically and compared with the code (\texttt{test\_lemma41}, \texttt{test\_derivation}); the exact heat exterior $E=c_\infty X^{-A}H(2d/X)$ is imposed as data and reproduced (\texttt{test\_heat\_exterior}); and a manufactured solution, a chosen smooth $(F,U)$ substituted into the equations to produce a forcing which the solver then inverts, checks the full residual and the gauge (\texttt{test\_manufactured}). The rule we followed: no number enters the paper without one of these behind it or a resolution sequence, and every failed attempt is written up in \texttt{results/FINDINGS\_live.md} with its reason. Roache \cite{roache02} on manufactured solutions.

\paragraph{The axis Cauchy problem: series, filter, march.} From analytic axis data the profiles are expanded in powers of $X$; the recursion \eqref{eq:R1}--\eqref{eq:R2} gives the coefficients, each a Chebyshev expansion in $\eta$ (\texttt{axis\_series.series\_coeffs}, \texttt{cheb\_projector}, \texttt{cheb\_spectrum}, \texttt{radius\_estimates} for the radius of convergence). Beyond that radius the state is marched in $X$ by an explicit high-order integrator, DOP853 with tight tolerances (\texttt{axis\_march.march}, \texttt{MarchCore}). The problem is ill posed in $X$: Chebyshev mode $m$ grows like $\exp\sqrt{8cmX}$ (\texttt{growth\_monitor} measures it), so the state is projected on the first $m_{\rm filter}\le28$ modes at every evaluation; this is a spectral filter in the sense of Vandeven \cite{vandeven91}, used as Hou and Li \cite{houli07} use it for nearly singular solutions, and the choice of $m_{\rm filter}$ is a resolution rule, not a tolerance. Hairer, N{\o}rsett and Wanner \cite{hairer93} for DOP853.

\paragraph{Fixed points for the pressure datum.} The datum is determined by matching the marched core to the exterior, a fixed-point problem $\Pi_0=G(\Pi_0)$ whose every evaluation of $G$ is a march. It is solved by Anderson acceleration, a least-squares mixing of the last few iterates (\texttt{axis\_join.anderson}), and, when that stalls, by Newton with a finite-difference Jacobian kept up to date by Broyden rank-one updates (\texttt{newton\_fixed\_point}). The datum has a fold (Section~\ref{sec:axis}, \texttt{axis\_pi0\_fold.py}), beyond which no fixed point exists; the fixed point iteration signals this by stalling, Newton by failing, and the fold is located by bracketing. Anderson \cite{anderson65}, Walker and Ni \cite{walker11}, Broyden \cite{broyden65}.

\paragraph{Matching and cone as least squares.} The moment identities of Theorem~4.6(v) of \OAI{} and the cone condition are imposed on a parametrized family of axis data and annulus content (\texttt{axis\_cone.Family}, \texttt{axis\_data\_from\_params}, \texttt{annulus\_from\_params}) by nonlinear least squares in the trust-region reflective form of SciPy (\texttt{match\_axis\_data}, \texttt{cone\_penalty}, \texttt{relaxed\_cone\_U}). Each residual evaluation is a full core: series, march, join. The pitfalls of Section~\ref{sec:axis} were all in the penalty: a capped penalty has no gradient on its plateau, cone quantities diverge where $F\to0$, and $a\ge2$ is impossible near the outer edge, so the penalty is based on $v_s\ge2$ with tapered margins. Branch, Coleman and Li \cite{branch99}.

\paragraph{Dynamic rescaling, BDF2 and the generalized eigenproblem.} To ask whether a self-similar profile attracts, one rewrites the equations in the frame that collapses with it, so that the profile is a steady state and the collapse time becomes $s=-\ln q$; the steady problem's residual is unchanged and a mass matrix marks which equations gain a time derivative (\texttt{annulus\_stability.mass\_matrix}). The linear stability is the generalized eigenproblem $\sigma Mv=-Jv$ with the Jacobian already in hand, solved by the QZ algorithm through SciPy (\texttt{spectrum}), and the nonlinear evolution by the second-order backward-difference formula with one Jacobian per step, step halving on failure and a restart by implicit Euler (\texttt{march}); a march along an eigenvector must reproduce its eigenvalue, which is the test (\texttt{tests\_stability.py}). States are saved with their grid and restarted on another grid by barycentric interpolation (\texttt{save\_state}, \texttt{interp\_state}), which turns an hour of continuation into a minute of Newton. McLaughlin et al.\ \cite{mclaughlin86} introduced dynamic rescaling for blow-up; Hairer and Wanner \cite{hairer96} for BDF; Moler and Stewart \cite{moler73} for QZ; Berrut and Trefethen \cite{berrut04} for barycentric interpolation.

\paragraph{Running it.} Python~3.10 with NumPy, SciPy, SymPy and Matplotlib; no compiled code. \texttt{README.md} in the repository gives the reference case and the tests; the run lists in \texttt{cluster/} are the exact command lines of every cluster job in this paper, and \texttt{HANDOFF.md} tells the story in the order it happened, including what went wrong.

\end{document}